\documentclass[aps,twocolumn,prd,showpacs,showkeys,preprintnumbers,superscriptaddress,nobibnotes,floatfix,longbibliography,notitlepage,nofootinbib]{revtex4-2}

\usepackage{amsmath}
\usepackage{amsfonts}
\usepackage{amssymb}
\usepackage{graphicx}
\usepackage[colorlinks=true,allcolors=blue]{hyperref}
\usepackage{relsize}
\usepackage{gensymb}
\usepackage{xspace}
\usepackage{mathptmx}
\usepackage{mathtools}
\usepackage{amsmath}
\usepackage{fontawesome}
\usepackage{placeins}
\usepackage[capitalize]{cleveref}
\usepackage[dvipsnames,table]{xcolor}
\usepackage[caption=false]{subfig}

\usepackage{siunitx}

\providecommand{\ehadreco}{\ensuremath{E_{\mathrm{had}}^{\mathrm{reco}}}\xspace}

\providecommand{\qz}{\ensuremath{q_{0}}\xspace}
\providecommand{\lownu}{low-$\nu$\xspace}

\providecommand{\enutrue}{\ensuremath{E_{\nu}^{\mathrm{true}}}\xspace}
\providecommand{\enureco}{\ensuremath{E_{\nu}^{\mathrm{reco}}}\xspace}

\providecommand{\nubar}{\ensuremath{\bar{\nu}}\xspace}
\providecommand{\nue}{\ensuremath{\nu_{e}}\xspace}

\providecommand{\numu}{\ensuremath{\nu_{\mu}}\xspace}
\providecommand{\numub}{\ensuremath{\nubar_{\mu}}\xspace}
\providecommand{\nutau}{\ensuremath{\nu_{\tau}}\xspace}

\providecommand{\argon}{$^{40}$Ar\xspace}
\providecommand{\tungsten}{$^{184}$W\xspace}
\providecommand{\faser}{FASER$\nu$2\xspace}
\providecommand{\flare}{FLArE10\xspace}
\def\bracketbar{\hbox{\kern-7pt\raise2pt%
    \hbox{{\tiny(}{\lower1.5pt\hbox{\bf--}}{\tiny)}}}}

\providecommand{\varpm}{\mathbin{\vcenter{\hbox{%
  \oalign{\hfil$\scriptstyle+$\hfil\cr
          \noalign{\kern-.3ex}
          $\scriptscriptstyle({-})$\cr}%
}}}}
\providecommand{\varmp}{\mathbin{\vcenter{\hbox{%
  \oalign{$\scriptstyle({+})$\cr
          \noalign{\kern-.3ex}
          \hfil$\scriptscriptstyle-$\hfil\cr}%
}}}}

\begin{document}

\title{A tolerable candle: the low-$\nu$ method with LHC neutrinos}

\author{C.~Wilkinson}
\email{cwilkinson@lbl.gov}
\affiliation{Lawrence Berkeley National Laboratory, Berkeley, CA 94720, USA}

\author{A.~Garcia~Soto}
\email{alfonsogarciasoto@fas.harvard.edu}
\affiliation{Department of Physics \& Laboratory for Particle Physics and Cosmology, Harvard University, Cambridge, MA 02138, USA}
\affiliation{Instituto de F{\'\i}sica Corpuscular (IFIC), CSIC and Universitat de Val{\`e}ncia, 46980 Paterna, Val{\`e}ncia, Spain}
\date{\today}

\begin{abstract}
  The Forward Physics Facility (FPF) plans to use neutrinos produced at the Large Hadron Collider (LHC) to make a variety of measurements at previously unexplored TeV energies. Its primary goals include precision measurements of the neutrino cross section and using the measured neutrino flux both to uncover information about far-forward hadron production and to search for various beyond standard model scenarios.
However, these goals have the potential to conflict: extracting information about the flux or cross section relies upon an assumption about the other.
In this manuscript, we demonstrate that the FPF can use the \lownu method --- a technique for constraining the flux shape by isolating neutrino interactions with low energy transfer to the nucleus --- to break this degeneracy.
We show that the \lownu method is effective for extracting the \numu flux shape, in a model-independent way. We discuss its application for extracting the \numub flux shape, but find that this is significantly more model dependent.
Finally, we explore the precision to which the \numu flux shape could be constrained at the FPF, for a variety of proposed detector options. We find that the precision would be sufficient to discriminate between various realistic flux models.
\end{abstract}

\maketitle

\section{Introduction}
Collisions at the Large Hadron Collider (LHC) produce a wealth of information, but existing LHC experiments characterize proton collisions with high transverse momentum, and have limited acceptance for particles produced roughly parallel to the beamline. Additional experiments are needed to probe the far-forward region, and sample the large flux of particles that are not accessible at existing experiments, including 0.1--10 TeV neutrinos of all flavors, which arise from the weak decay of hadrons with large rapidity produced in proton-proton collisions~\cite{DeRujula:1992sn}. These experiments have a broad physics program that encompasses both standard model physics and beyond-standard model searches.
The FASER and SND@LHC experiments are currently operating experiments in the far-forward region, and during LHC Run 3 have detected LHC neutrinos for the first time~\cite{FASER:2023zcr,FASER:2021mtu,PhysRevLett.131.031802} --- a major achievement in Particle Physics.
During Run 3, the FASER and SND@LHC experiments expect to have $\mathcal{O}$(10k) and $\mathcal{O}$(1k) neutrino charged-current (CC) interactions in their instrumented volume~\cite{FASER:2019dxq,SNDLHC:2022ihg}.
Additionally, the planned Forward Physics Facility (FPF)~\cite{Anchordoqui:2021ghd,Feng:2022inv} is intended to operate in the high-luminosity LHC era, and will collect orders of magnitude more neutrino interactions during its planned run with an integrated luminosity of 3000 fb$^{-1}$.
The neutrino physics goals of the FPF include tests of lepton flavor universality, measurements of the cross-sections of all three flavors in a region not previously measured before, tests of quantum chromodynamics (QCD) and searches for beyond-standard-model (BSM) effects that may show up in their production, propagation or interaction.

\begin{figure}[htbp]
  \centering
  \captionsetup[subfloat]{captionskip=-1pt}
  \includegraphics[width=0.9\linewidth]{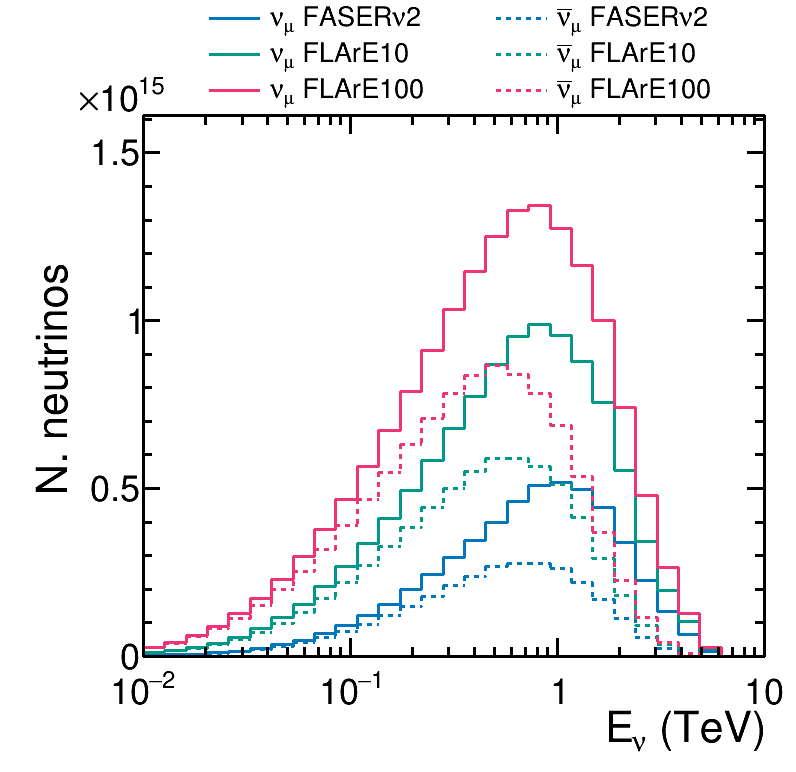}
  \caption{A comparison of the predicted \numu and \numub fluxes that would be sampled by three candidate FPF detectors, \faser, FLArE10 and FLArE100, assuming an integrated luminosity of 3000 fb$^{-1}$. Reproduced from Ref.~\cite{Kling:2021gos} (using the SIBYLL v2.3d model~\cite{Riehn:2019jet} for both light and charmed hadron production).}
  \label{fig:flux_comp}
\end{figure}
The neutrino flux at the FPF will be dominated by $\numu^{\bracketbar}$, with an order of magnitude fewer $\nue^{\bracketbar}$'s and three orders of magnitude fewer $\nutau^{\bracketbar}$'s, the latter being suppressed due to the high mass of their charged partner. Figure~\ref{fig:flux_comp} shows the expected $\numu^{\bracketbar}$ fluxes for three candidate FPF detectors (described in detail in Section~\ref{sec:reco}), \faser, FLArE10 and FLArE100, which have different dimensions transverse to the incoming proton beam, and sample different fluxes as a result.
The fluxes shown in Figure~\ref{fig:flux_comp} and used throughout this work were provided by the authors of Ref.~\cite{Kling:2021gos}. They are produced using a variety of hadron production models as implemented in the CRMC simulation package~\cite{ulrich_ralf_2021_5270381}, to simulate primary collisions; then a fast neutrino flux simulation described in Ref.~\cite{Kling:2021gos} is used to simulate their propagation through the LHC beam pipe and magnets and their decays into neutrinos. Figure~\ref{fig:flux_comp} uses the SIBYLL v2.3d model~\cite{Riehn:2019jet} for both light and charmed hadron production in the primary collisions.
For each candidate detector, both the \numu and \numub fluxes peak at $\approx$1 TeV, and the bulk of the flux distribution falls in the region 0.1--4 TeV.

At the FPF, as well as FASER and SND@LHC, the $\numu^{\bracketbar}$ flux below 100 GeV is dominated by pion decay, whereas kaon and charmed hadron decays are the leading contributors at higher energies. However, precise estimation of these rates is currently challenging due to our limited knowledge of hadronic yields in the forward regime~\cite{Bai:2020ukz,Bai:2021ira,Bai:2022jcs,Kling:2021gos,Maciula:2022lzk,Buonocore:2023kna,Bhattacharya:2023zei}. Differences between flux models go from tens of percent below 100 GeV to factors of up to 10 at a few TeV, where the contribution from charmed hadron decays dominates. Various BSM scenarios may also radically alter the neutrino flux distribution that will be sampled by far-forward detectors. Neutrino flux measurements at these detectors would therefore provide a tool for disambiguating between and constraining various standard model and BSM production processes.

The 0.1--10 TeV neutrino cross section falls between measurements made at lower energies with accelerator neutrino beams~\cite{zeller12, Katori:2016yel, NuSTEC:2017hzk, Mahn:2018mai}, and those made at $\gtrsim 10$ TeV energies with neutrino telescopes using neutrinos from astrophysical sources~\cite{IceCube:2017roe,Bustamante:2017xuy,IceCube:2017roe}. Measurements in this region by the FPF would fill this gap, and be particularly impactful for understanding the prompt atmospheric neutrino flux~\cite{Bhattacharya:2016jce,Bhattacharya:2015jpa,Gauld:2015kvh,Ostapchenko:2022thy,Zenaiev:2019ktw,Benzke:2017yjn,Garzelli:2016xmx,Enberg:2008te}, an as yet undetected but guaranteed source of high-energy neutrinos, which suffers from large uncertainties in the modeling of heavy hadrons. A precise estimation of this component is crucial for the characterization of astrophysical neutrinos. Additionally, the observation of muon multiplicities in extensive air showers generated by cosmic rays cannot be adequately explained by state-of-the-art predictions~\cite{EAS-MSU:2019kmv}. Several studies indicate this discrepancy arises from unaccounted physical effects governing soft hadronic interactions at high energies~\cite{Albrecht:2021cxw}.

Measurements of the neutrino flux and the neutrino cross section at the FPF are both important. However, both are currently unknown, which presents a problem. Only the interaction rate can be measured directly, which is the convolution of the flux and the cross section (along with the detector efficiency). Making a measurement of either requires assumptions to be made about the other. However, various methods have been used by past experiments to characterize the flux by using subsamples with known cross sections to break this degeneracy. For example, neutrino-electron elastic scattering and inverse muon decay have both been used as ``standard candles'' by few-GeV accelerator neutrino experiments~\cite{Tomalak:2019ibg,MINERvA:2015nqi,MINERvA:2019hhc,Marshall:2019vdy}. However, while their cross sections can be calculated, they are very small, and are unlikely to be useful for the FPF --- but are for higher intensity, lower energy experiments. Additionally, the ``\lownu method'', in which a sample of neutrino interactions with low energy transfer to the nucleus are isolated, have been used at high-energy accelerator neutrino experiments~\cite{Belusevic:1987rn,Belusevic:1988ab,Auchincloss:1990tu,Seligman:1997fe,Mishra:1990ax}, but has been shown to be model-dependent at lower energies~\cite{Wilkinson:2022dyx}. In this paper, we explore the possibility of utilizing this method at the FPF.

The remaining sections of the manuscript are structured as follows. Section~\ref{sec:lownu} introduces the \lownu method. Section~\ref{sec:true} evaluates the applicability and validity of the ``\lownu'' method within the energy range relevant to the FPF facility. Section~\ref{sec:reco} focuses on the analysis of the FPF detectors' capability to effectively isolate a sufficiently large sample of events for the application of this method. Section~\ref{sec:fit} provides an estimation of the constraints that can be imposed on the neutrino flux through an example analysis. Finally, we present our conclusions in Section~\ref{sec:conclusions}.

\section{The \lownu method}
\label{sec:lownu}
The \lownu method refers to a sample in which the energy transfer to the nucleus is small. It was first developed for the CCFR experiment~\cite{Auchincloss:1990tu,Seligman:1997fe}, is generally attributed to Ref.~\cite{Mishra:1990ax}, although is closely related to prior work described in Refs.~\cite{Belusevic:1987rn,Belusevic:1988ab}. A history of the use of the \lownu method has been collated in Ref.~\cite{Wilkinson:2022dyx}.

For charged-current interactions,  $\nu \equiv \qz=E_{\nu}-E_{l}$, where $E_{\nu}$ is the incoming neutrino energy, and $E_{l}$ is the energy of the outgoing charged lepton. As $\nu$ is overloaded in neutrino physics, here we will follow the convention of Ref.~\cite{Wilkinson:2022dyx} and use ``\qz'' to denote the energy transfer and ``\lownu'' to denote the method. The \lownu method is motivated by the expression of the inclusive charged-current scattering cross section commonly used in deep inelastic scattering (DIS) theory, written in terms of nucleon structure functions. The cross section, differential in \qz, is found by integrating $d^2\sigma/d\qz dx$ over $x = Q^2/(2M \qz)$,
\begin{align}
  \frac{\mathrm{d}\sigma}{\mathrm{d}\qz} &= \frac{G^2_{\mathrm{F}} M}{\pi} \int_0^1 \Big(F_2 - \frac{\qz}{E_\nu}  \left[F_2 \varpm xF_3\right]
      + \frac{\qz}{2E_\nu^2}  \left[\frac{Mx(1-R_{\mathrm{L}})}{1+R_{\mathrm{L}}}F_2\right]  \nonumber \\
      &+ \frac{\qz^2}{2E_\nu^2}  \left[\frac{F_2}{1+R_{\mathrm{L}}} \varpm xF_3\right]  \Big)\,\mathrm{d}x,
\label{eq:low-nu}
\end{align}
\noindent where $M$ is the struck nucleon mass, $F_2$ and $xF_3$ are structure functions and $R_{\mathrm{L}}$ is the structure function ratio $F_2/(2xF_1)$, with $G_{\mathrm{F}}$ being Fermi's constant, and the $+(-)$ is used for (anti-)neutrinos~\cite{Mishra:1990ax}.

The key idea behind the \lownu method is that for low values of \qz, the $\qz/E_{\nu}$ terms in Equation~\ref{eq:low-nu} are small, so the cross section is approximately constant with $E_{\nu}$. If a sample of events with low \qz can be isolated experimentally, it can be used to measure the flux {\it shape} as a function of the neutrino energy. The method relies on three key requirements:
\begin{enumerate}
\item There is a region in true \qz which is approximately constant as a function of neutrino energy.
\item This sample be selected in FPF detectors without introducing significant model dependence.
\item This region produces a usefully large sample of events in the FPF detectors
\end{enumerate}

In Ref.~\cite{Wilkinson:2022dyx}, it was argued that the \lownu method is model-dependent at few-GeV energies, where the \qz values of interest are sub-GeV, and the DIS formalism breaks down. At low \qz ($\lesssim 2$ GeV), the neutrino-nucleus cross section is dominated by non-DIS processes, largely quasielastic scattering, and resonance pion production, which are not well described by Equation~\ref{eq:low-nu}. However, although this is problematic in the few-GeV neutrino case, these contributions to the cross section are saturated (and therefore constant) for $\enutrue \gtrsim 20$ GeV~\cite{Belusevic:1988ab}. In the high-energy FPF regime, these contributions will not vary with energy and do not necessarily compromise the \lownu method. Indeed, the \lownu method has been used successfully in both CCFR and NuTeV, with neutrino energies in the range $30 \leq \enutrue \leq 360$ GeV, which both used \lownu samples with $\qz \leq 20$ GeV. However, in order to assess the impact that the $\qz/E_{\nu}$ and $\qz^{2}/E_{\nu}^{2}$ terms in Equation~\ref{eq:low-nu} have on the \lownu sample, using data-driven corrections, they had to exclude the non-DIS contributions by further to CCFR (NuTeV) further restricted the sample of interest used by CCFR (NuTeV) to $4 \leq \qz \leq 20$ GeV ($5 \leq \qz \leq 20$ GeV).

It is also important that the requirements that should be placed on the \lownu method for the FPF is radically different to the few-GeV accelerator neutrino case. In the latter, there are 5--10\% flux shape uncertainties, already well constrained by replica target experiments. In this case, a few-percent bias in the \lownu method would be large relative to the prior uncertainty, so must be well understood to be a reasonable trade-off. For the FPF, the standard model flux uncertainty is significantly larger, and large changes to the flux are part of the BSM program. The \lownu method would be likely to provide a useful tool for breaking the neutrino flux and cross section degeneracy even at the cost of a small model-dependent bias.

\section{Neutrino cross-section model}
\label{sec:model}

Our understanding of neutrino--nucleus scattering across the 0.1--10 TeV energies relevant for the FPF is limited~\cite{Bodek:2021bde,Candido:2023utz,Xie:2023suk,Jeong:2023hwe}, lacking both experimental data and consistent theoretical predictions.
At these energies, the dominant interaction mechanism is deep inelastic scattering (DIS), for which there have been a variety of measurements made, albeit at lower neutrino energies, $\mathcal{O}$(10--100 GeV)~\cite{deGroot:1978feq,Mukhin:1979bd,GargamelleSPS:1981hpd,Berge:1987zw,Anikeev:1995dj,NuTeV:2005wsg,NOMAD:2007krq,MINOS:2009ugl}.
These measurements typically used heavy nuclei as targets, and various nuclear effects had to be considered to describe the available data accurately. However, the current theoretical frameworks available are unable to explain all of the observed neutrino data~\cite{Muzakka:2022wey}. Various hypotheses have been proposed to explain the discrepancies, including the possibility of an alternative mechanism for shadowing in neutrino--nucleus interactions or potential issues in the acquisition of experimental data.

In this work, we use two different GENIE models, which have been tuned to low-energy and high-energy data respectively~\cite{GENIE:2021npt} (from now on, we will refer to them as LE and HE), to model neutrino interactions in the FPF detectors. The GENIE outputs are minimally processed with NUISANCE~\cite{Stowell:2016jfr}. As previously described, DIS is the dominant process in this energy regime. The critical input for modeling DIS interactions is the description of the nucleon structure functions, so in the next paragraphs, we will briefly describe their implementation in both LE and HE models.

The GENIE LE model uses the Bodek-Yang prescription~\cite{Bodek:2004pc}, which is commonly used in simulations developed for the long-baseline neutrino oscillation community~\cite{Ruso:2022qes}. This model provides a phenomenological description of the structure functions. Particularly relevant is the implementation for $Q^2<1$ GeV$^2$, where perturbative QCD breaks down (this contribution can be up 20\% when $\enutrue \leq 20$ GeV). In their approach, leading order expressions for the structure functions were modified, including a Nachtmann scaling variable~\cite{Nachtmann:1973mr}. They also multiply all parton density functions (PDFs) by $Q^2$-dependent terms (so called, K factors~\cite{CCFR:2000ihu}). These parameters account for several effects: dynamic higher twist, higher-order QCD terms, transverse momentum of the initial quark, the effective masses of the initial and final quarks originating from multi-gluon interactions at low-$Q^2$, and the correct form in the low-$Q^2$ photo-production limit. The parameters were extracted from a fit to inelastic charged-lepton scattering data on hydrogen and deuterium targets~\cite{Whitlow:1991uw,BCDMS:1989ggw,NewMuon:1996fwh,H1:2003xoe} using GRV98LO as input PDFs~\cite{Gluck:1998xa}. The main limitation of the Bodek-Yang prescription is that it is not reliable at energies above a few TeV, where high-$Q^2$ interactions dominate. In this regime, the structure functions converge to the leading-order approximation with GRV98LO. It should be noted that at few-TeV energies, we are using the LE model outside the region where the model authors recommend its use, although the low-\qz region most of interest for this work is well within the region in which it should be valid.

The GENIE HE model was developed to describe the high-Q regime~\cite{Garcia:2020jwr}, in which the structure functions can be factorized in terms of coefficient functions and PDFs using perturbation theory. The PDFs are extracted from experimental data. The evolution of these PDFs is determined by the solutions of the DGLAP evolution equations~\cite{Gribov:1972ri,Dokshitzer:1977sg,Altarelli:1977zs}. The coefficient functions can be computed in perturbation theory as a power expansion in the strong coupling $\alpha_{\mathrm{s}}$. In this work, we adopted the CSMS model~\cite{Cooper-Sarkar:2011jtt} as the baseline because it has been an important benchmark for the neutrino telescope community. As inputs, this calculation uses the next-to-leading-order HERA1.5 PDF set~\cite{Cooper-Sarkar:2010yul} and coefficient functions from QCDNUM~\cite{Botje:2010ay}.

Bodek-Yang and CSMS models can be used to compute the kinematics of the outgoing leptons and quarks. In both LE and HE models, the subsequent hadronization of the partonic shower is carried out using PYTHIA6~\cite{Sjostrand:2006za}. Finally, the GENIE LE tune model final state interactions of the hadrons inside the nuclei using INTRANUKE~\cite{Andreopoulos:2015wxa}, while the HE tune neglects them.

\begin{figure}[htbp]
  \centering
  \captionsetup[subfloat]{captionskip=-1pt}
  \includegraphics[width=0.9\linewidth]{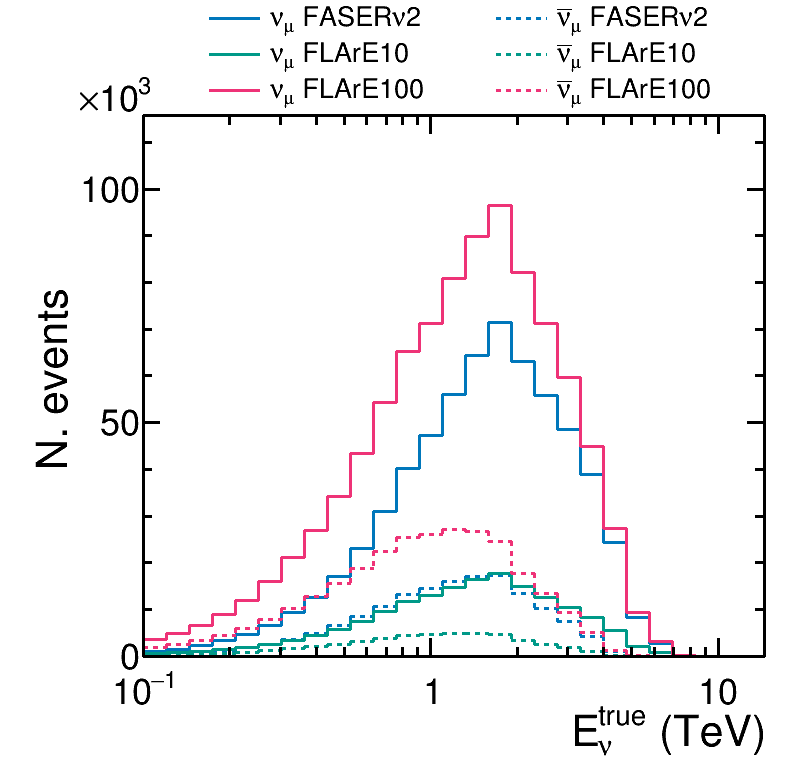}
  \caption{The total predicted \numu and \numub charged-current event rates expected for the three candidate FPF detectors, \faser (20 t \tungsten), FLArE10 (10 t \argon) and FLArE100 (100 t \argon), using the fluxes shown in Figure~\ref{fig:flux_comp} for an integrated luminosity of 3000 fb$^{-1}$ and the LE GENIE model.}
  \label{fig:total_rate_comp}
\end{figure}
Figure~\ref{fig:total_rate_comp} shows the total predicted charged-current event rates for both \numu and \numub fluxes at the three FPF detectors considered in this work, \faser (20 t \tungsten), FLArE10 (10 t \argon) and FLArE100 (100 t \argon), using the GENIE LE model and the fluxes taken from Ref.~\cite{Kling:2021gos} and shown in Figure~\ref{fig:flux_comp}. The cross section increases approximately linearly with \enutrue in this region. The predictions for the GENIE HE model are not shown, but are within $\approx$10--20\% of the LE model as a function of \enutrue.

\section{Low-$\nu$ performance at truth level}
\label{sec:true}
\begin{figure*}[htbp]
  \centering
  \captionsetup[subfloat]{captionskip=-1pt}
  \subfloat[\numu--\tungsten]  {\includegraphics[width=0.45\linewidth]{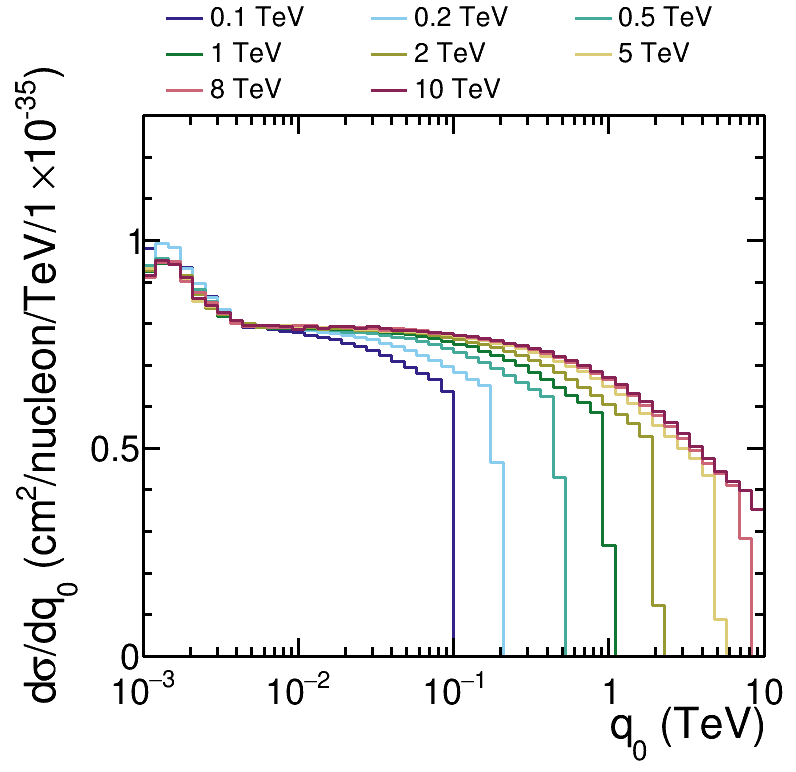}}
  \subfloat[\numu--\argon]     {\includegraphics[width=0.45\linewidth]{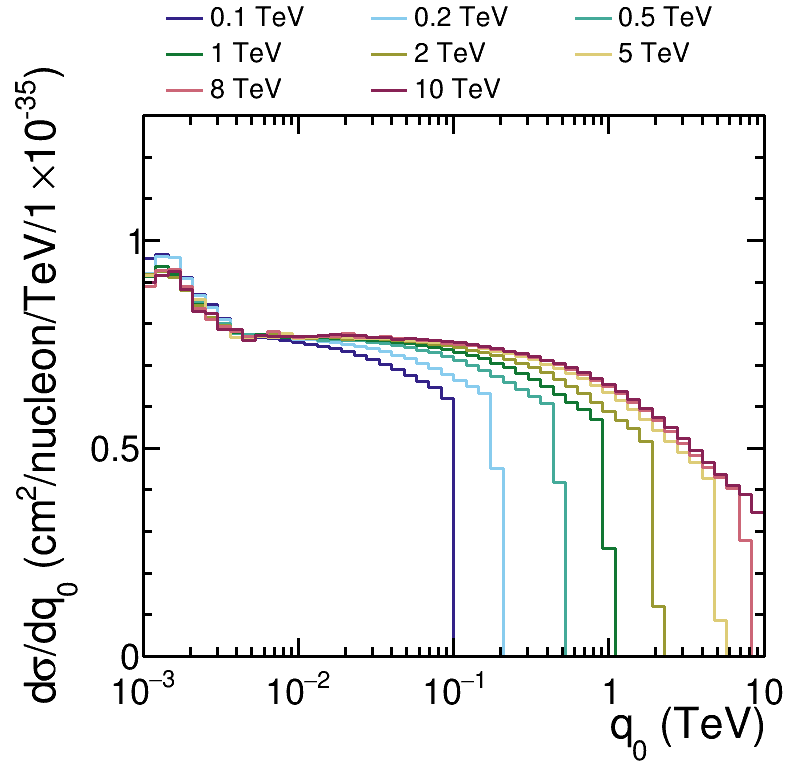}}\\\vspace{-10pt}
  \subfloat[\numub--\tungsten] {\includegraphics[width=0.45\linewidth]{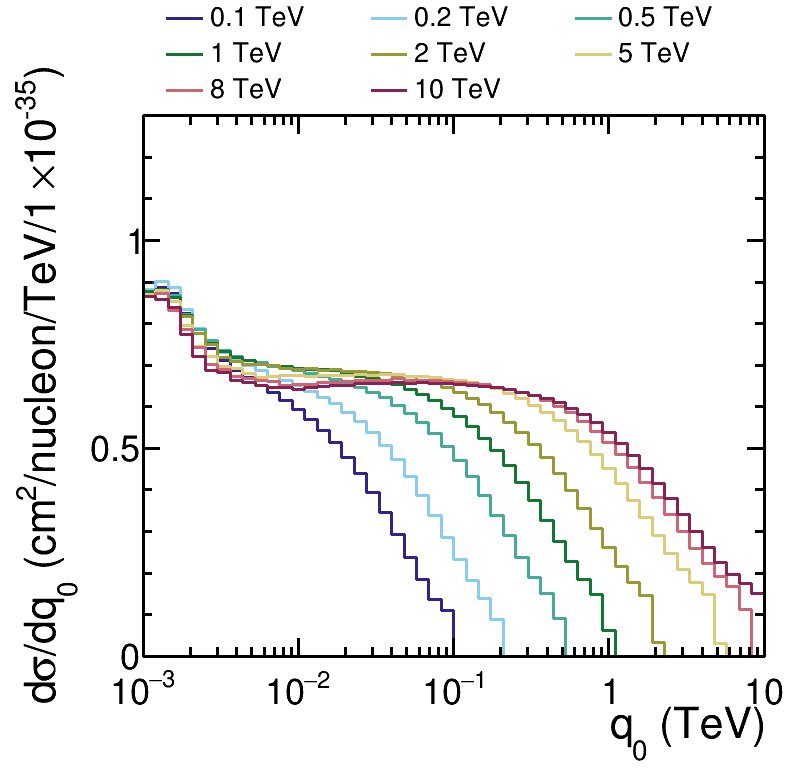}}
  \subfloat[\numub--\argon]    {\includegraphics[width=0.45\linewidth]{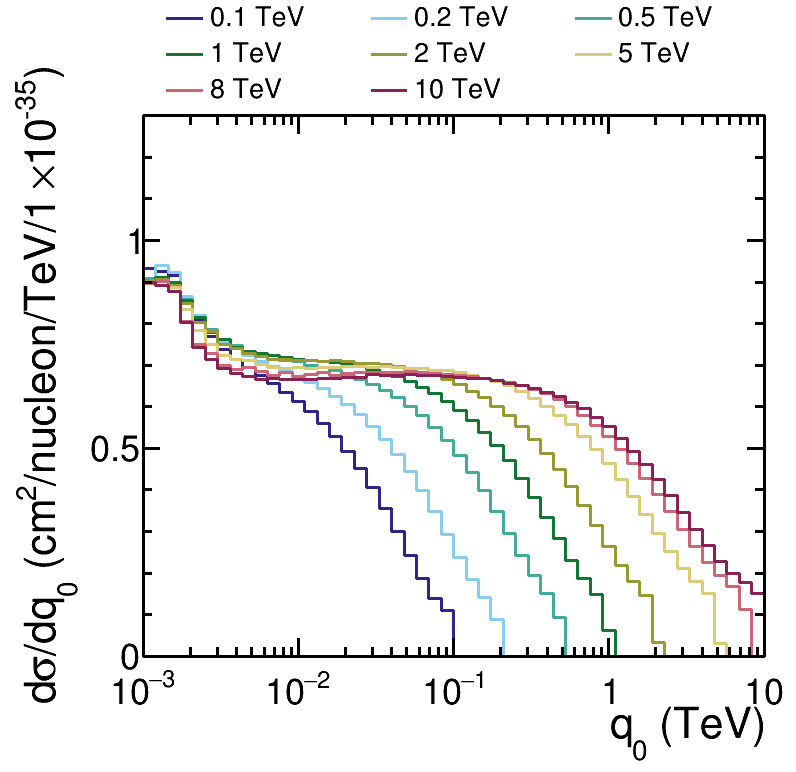}}
  \caption{The \numu (top) and \numub (bottom) per nucleon CC-inclusive cross sections for both \tungsten (left) and \argon (right) targets, using the GENIE LE tune, for fixed values of true neutrino energy, as a function of true \qz.}
  \label{fig:true_xsec_comp}
\end{figure*}
Figure~\ref{fig:true_xsec_comp} shows the \numu (top) and \numub (bottom) CC-inclusive cross sections for both \tungsten and \argon targets, using the GENIE LE tune, for fixed values of \enutrue, as a function of true \qz. For each \enutrue value, there is a small peak around $1 \leq \qz \leq 3$ GeV, which is due to overlapping resonant pion production below the DIS region. The resonant peak is more pronounced for \argon than \tungsten as both the \numu--proton and \numub--proton resonant pion production cross sections are larger than the \numu--neutron and \numub--neutron cross sections and the proton fraction is larger in \argon than \tungsten. At higher $\qz \gtrsim 5$ GeV, the rise is dominated by the DIS cross section turning on, which later falls. The behavior is qualitatively similar for \numu--\argon and \numu--\tungsten (as well as \numub--\argon and \numub--\tungsten) due to the similar DIS cross sections for neutrons and protons. The sharp cut-off at the highest \qz values for each \enutrue value is at a kinematic limit ($\qz < \enutrue$).

As may be expected given the basic premise of the \lownu method, the cross sections shown in Figure~\ref{fig:true_xsec_comp} are approximately constant with \enutrue at low values of \qz, but diverge at high-\qz. The divergence is faster for \numub than \numu, which can be readily understood by the sign changes in Equation~\ref{eq:low-nu}.
For \numu, the cross section seems to be appoximately constant as a function of \enutrue for $\qz \lesssim 20$ GeV, which is in keeping with the \lownu sample definitions used by CCFR and NuTeV.
Here, we also use this as the region of interest for defining a \lownu sample for \numu--\tungsten and \numu--\argon interactions at the FPF.
There are already $\approx$10\% differences in the \numub cross sections shown at $\qz = 20$ GeV, so we use a more restrictive cut of $\qz \leq 10$ GeV to define a \lownu sample for both \numub--\tungsten and \numub--\argon interactions.
There is a trade-off between increasing the sample size with a higher \qz cut, which would improve the eventual flux constraint, and reducing the potential model dependence by decreasing the \qz cut, which would need to be properly assessed for a real data analysis.
Neither of the cut values defined here are optimized, but are sufficiently well motivated for this preliminary study.

\begin{figure}[htbp]
  \centering
  \captionsetup[subfloat]{captionskip=-1pt}
  \subfloat[\numu--\tungsten, $E_{\nu}$ = 1 TeV]   {\includegraphics[width=0.9\linewidth]{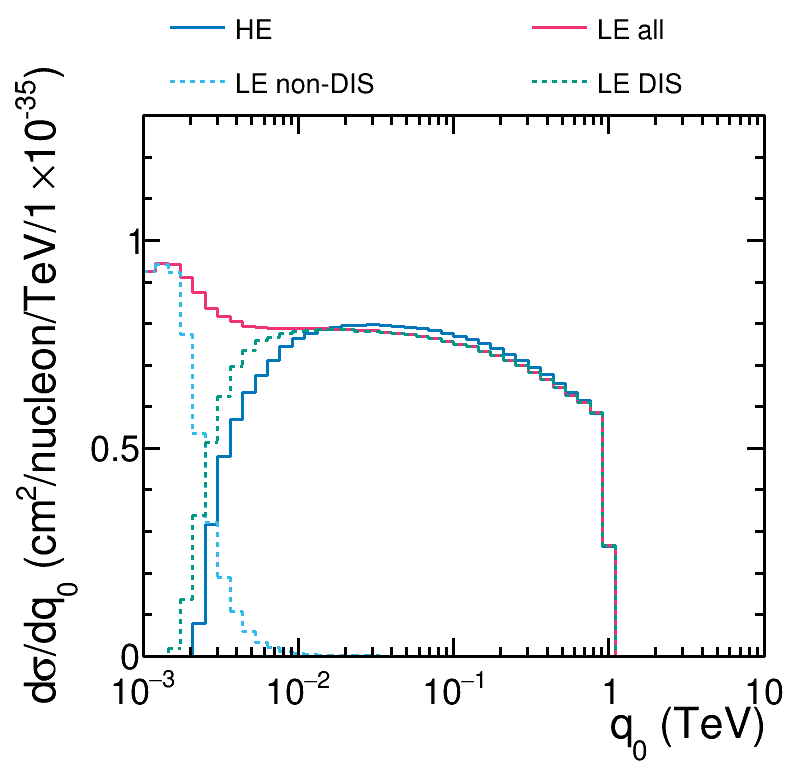}}\\\vspace{-10pt}
  \subfloat[\numu--\argon, $E_{\nu}$ = 1 TeV]      {\includegraphics[width=0.9\linewidth]{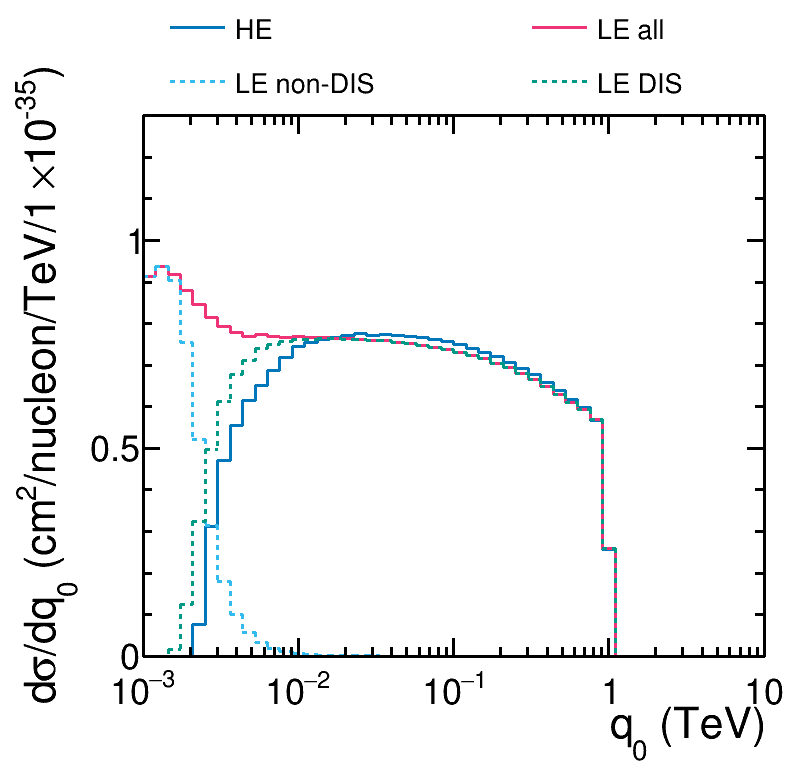}}
  \caption{Comparison of the \numu--\tungsten and \numu--\argon cross sections as a function of true \qz, for both GENIE LE and HE tunes at $\enutrue=1$ TeV. The LE tune contributions are shown broken down into DIS and non-DIS contributions. The HE tune only includes DIS contributions.}
  \label{fig:true_xsec_model_comp_numu}
\end{figure}
Figure~\ref{fig:true_xsec_model_comp_numu} shows the \numu--\tungsten and \numu--\argon cross sections as a function of true \qz, for both GENIE LE and HE tunes at $\enutrue=1$ TeV. The LE tune has additionally been split into DIS and non-DIS components, whereas the HE tune only includes DIS contributions. The general trends of Figure~\ref{fig:true_xsec_model_comp_numu} are qualitatively similar across the range of energies explored in this work, and for \numub--\tungsten and \numub--\argon interactions. The non-DIS contributions dominate the cross section for the LE tune for $\qz \lesssim 3$~GeV, remain significant up to $\qz \approx 5$ GeV, and are larger for \numu--\argon than \numu--\tungsten due to the larger proton fraction. The differences in the shape of the LE and HE DIS contributions differ between \argon and \tungsten.

\begin{figure}[htbp]
  \centering
  \captionsetup[subfloat]{captionskip=-1pt}
  \subfloat[\numu--\tungsten]  {\includegraphics[width=0.9\linewidth]{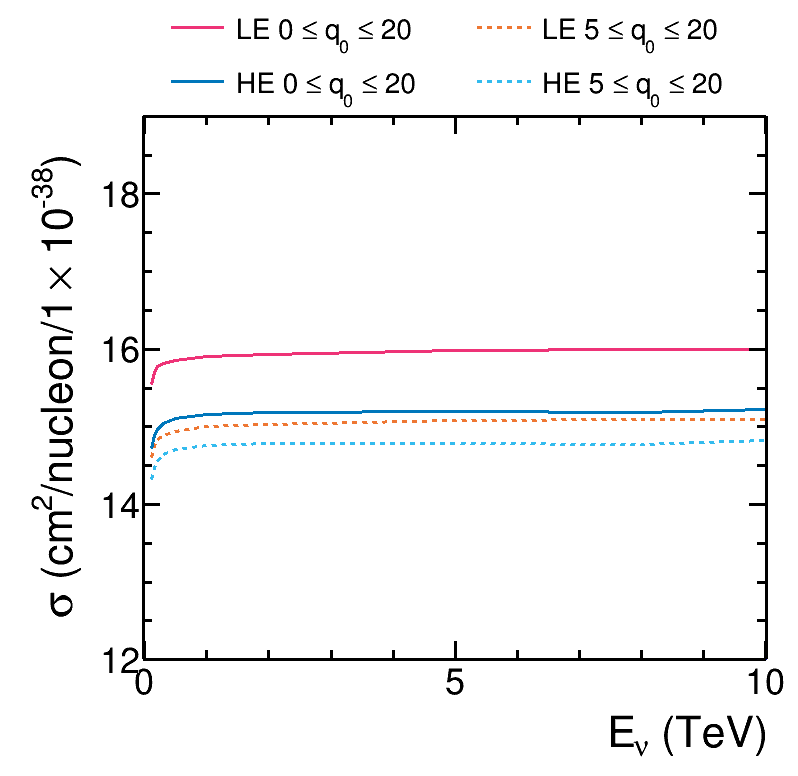}}\\\vspace{-10pt}
%  \subfloat[\numu--\argon]     {\includegraphics[width=0.45\linewidth]{figures/enu_corr_1000180400_14_q0_xsec.png}}\\\vspace{-10pt}
  \subfloat[\numub--\tungsten] {\includegraphics[width=0.9\linewidth]{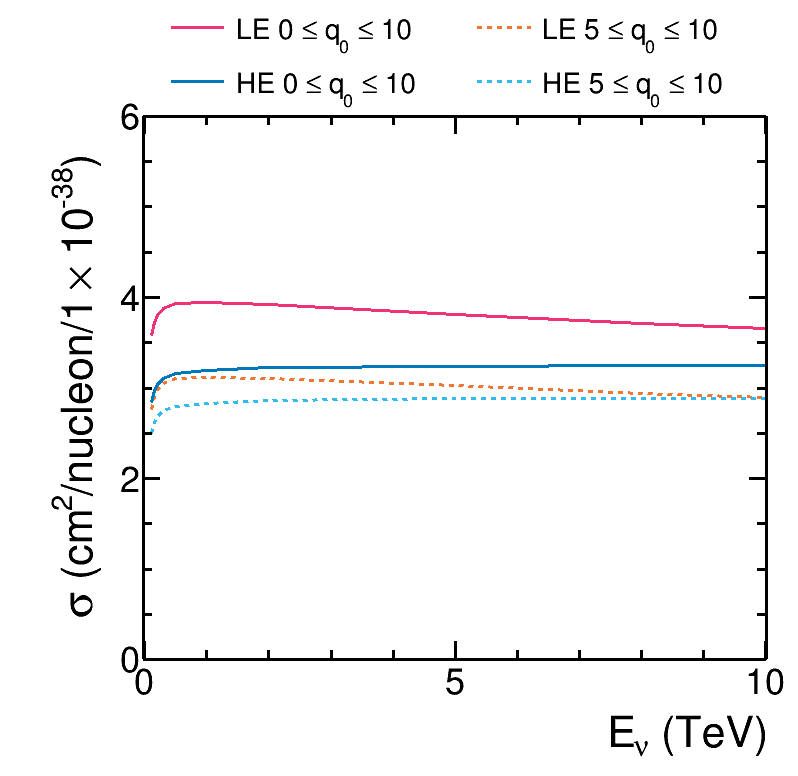}}
%  \subfloat[\numub--\argon]    {\includegraphics[width=0.45\linewidth]{figures/enu_corr_1000180400_-14_q0_xsec.png}}  
  \caption{The \numu (top) and \numub (bottom) restricted \lownu cross section for a \tungsten target, selected using cuts on true \qz, shown as a function of \enutrue for both GENIE LE and HE tunes.}
  \label{fig:lownu_q0_xsec}
\end{figure}  
Figure~\ref{fig:lownu_q0_xsec} shows the restricted \lownu cross section, defined here as the CC-inclusive cross section with a restriction of $\qz \leq 20$ GeV ($\qz \leq 10$ GeV) for \numu--\tungsten (\numub--\tungsten), as a function of \enutrue. The general behavior for \numu--\argon and \numub--\argon is very similar. For both \numu--\tungsten and \numu--\argon, the cross section is relatively flat across the energy range of interest (0.1--10 TeV), with $\approx$3\% differences at the lowest energies investigated here. The GENIE LE and HE models have a different cross section, as may be expected from Figure~\ref{fig:true_xsec_model_comp_numu}, but despite having very different DIS predictions they have a very similar {\it shape}, which is the important feature for the \lownu method. For both \numub--\tungsten and \numub--\argon, the cross section is also relatively flat, although the LE and HE models have markedly different shapes, and both have large $\approx$10\% differences across the energy range of interest, indicating that the method is likely to perform significantly less well for \numub than \numu.

Alternative \lownu cross sections are also shown in Figure~\ref{fig:lownu_q0_xsec} with an additional minimum \qz restriction of $\qz \geq 5$ GeV. This is intended to cut out the non-DIS portion of the cross section, for which Equation~\ref{eq:low-nu} does not apply. However, for the neutrino energies of interest, the quasielastic-like and resonance channels which dominate this region are likely to be fully saturated and energy independent, so it may not be necessary to remove this region, depending on the analysis approach taken, as discussed in Section~\ref{sec:lownu}. As expected, this additional restriction has a larger impact on the LE than HE tune, the latter of which only includes DIS interactions. The $\qz \geq 5$ GeV restriction has a relatively small contribution to the total \lownu cross section for \numu at $\approx$10\%, but a larger one for \numub at $\approx$25\%, which is unsurprising given the smaller \qz range used for \numub. The energy independence of all of the cross sections is very similar with and without the additional $\qz \leq 5$ GeV restriction included.  

Figure~\ref{fig:lownu_q0_xsec} demonstrates that the first requirement for the \lownu method to work, as identified in Section~\ref{sec:lownu}, is fulfilled for \numu interactions, and possibly for \numub interactions --- there is a region in true \qz which is approximately constant as a function of neutrino energy.

\FloatBarrier
\section{Impact of detector thresholds and reconstruction}
\label{sec:reco}
Although the \lownu method refers to the true energy transfer, \qz, this is not experimentally accessible --- detectors have thresholds and may not be able to reconstruct some particles altogether (e.g., neutrons, or outgoing neutrinos), or may not be able to associate them to the neutrino interaction reliably (e.g., $K^0_{\mathrm{L}}$'s). Energy or momentum reconstruction of particles also has some uncertainty due to the resolution of the detector. Similarly, the true neutrino energy, \enutrue, is not directly accessible. Both \enutrue and \qz must be reconstructed using detector observable quantities. Additionally, in high-background environments, such as is the case for the FPF, there may also be selection limitations that affect the reconstruction efficiency as a function of \qz, such as a minimum number of tracks needed to identify a vertex. 

\begin{table}[htbp]
  {\renewcommand{\arraystretch}{1.2}
    \begin{tabular}{c|ccc}
      \hline\hline
      & \faser & FLArE10 & FLArE100\\
      \hline
      Fiducial mass      & 20 t        & 10 t  & 100 t \\
      Det. cross-section & 0.5$\times$0.5 m & 1.0$\times$1.0 m & 1.6$\times$1.6 m \\
      Target material    & $^{184}$W    & \multicolumn{2}{c}{$^{40}$Ar} \\
      Muon resolution    & 5\%         & \multicolumn{2}{c}{5\%}  \\
      Charged had. res.   & 50\%        & \multicolumn{2}{c}{30\%} \\
      Charged had. threshold & $p \geq 300$ MeV & \multicolumn{2}{c}{$p \geq 30$ MeV} \\
      EM shower res.     & 50\%        & \multicolumn{2}{c}{30\%} \\
      Minimum track cut  & 5           & \multicolumn{2}{c}{N/A}  \\
      Invisible particles& \multicolumn{3}{c}{$n$, $\bar{n}$, $K^0_{\mathrm{L}}$, $\nu_{X}$} \\
      \hline\hline
    \end{tabular}
  }
  \caption{Assumptions used for the various FPF detector options considered in this work, based on Refs.~\cite{FASER:2019dxq, Anchordoqui:2021ghd, Feng:2022inv,Batell:2021blf,DUNE:2020jqi}.}
  \label{tab:detector_model}
\end{table}
Tab.~\ref{tab:detector_model} summarizes the models used to approximate the detector response for both \faser and FLArE throughout this work, using numbers taken from Refs.~\cite{FASER:2019dxq, Anchordoqui:2021ghd, Feng:2022inv}.
Details of the detector cross section and mass are taken from Ref.~\cite{Feng:2022inv}.
For simplicity, we neglect the small contribution from materials other than \tungsten in \faser.
Detailed information about the neutrino flux distributions for each detector is provided by Ref.~\cite{Kling:2021gos}, assuming an integrated luminosity of 3000 fb$^{-1}$.
A 5\% momentum resolution on muons has been assumed, following Ref.~\cite{Feng:2022inv}, and assuming that muons are reconstructed using the magnetized FASER detector downstream of each of these proposed detector components.
Additionally, to simplify the analysis, we assume perfect sign selection for muons, although this may break down at the highest energies.
The charged hadron and EM shower energy resolutions of 50\% for \faser, motivated by Ref.~\cite{FASER:2019dxq}, and apply a $p \geq 300$ MeV threshold for all charged particles, motivated by discussion in Ref.~\cite{Batell:2021blf}.
We also follow the description in Ref.~\cite{FASER:2019dxq} and apply a minimum track cut for the \faser detector, which is required in order to unambiguously identify a vertex when the emulsion films are scanned offline after taking a reasonably large exposure. Note that the track cut does not include any EM contributions, and requires $\geq 5$ charged hadrons that are above detection threshold.
We take the DUNE hadronic energy uncertainty of $\approx$30\% (see, for example, Ref.~\cite{DUNE:2020jqi}) as a motivation for both charged hadron and EM shower resolutions for FLArE, and apply a $p \geq 30$ MeV threshold for all charged particles, motivated by discussion in Ref.~\cite{Batell:2021blf}.
We do not apply a minimum track cut for FLArE, assuming that its relatively fast timing will make it easier to separate beam related events, although the high muon flux at the FPF may make it necessary to include some additional activity or minimum track cut to separate neutrino events from misreconstructed muon backgrounds.
In reality, the resolutions for both \faser and FLArE will depend on the particle momenta, but are constant in our analysis, which may make them overly conservative.
Additionally, we assume that neutrons (both $n$ and $\bar{n}$), $K^0_{\mathrm{L}}$'s, and neutrinos are simply not reconstructed for both detectors.

We note that the detector model used here is necessarily naive.
In particular, we acknowledge that some FPF studies have used much smaller estimates for particle or neutrino energy resolutions.
Ref.~\cite{FASER:2019dxq} describes an approach to obtain a neutrino energy resolution of $\approx$30\%, using angular information to constrain the event reconstruction with a neural network, which would imply better hadronic and EM shower energy resolution than we assume, but this approach may lead to model dependence, particularly in corners of phase space such as are relied upon by the \lownu method.
Additionally, studies in Ref.~\cite{Feng:2022inv} consider sensitivities with both a 15\% and 45\% charged particle energy resolution, which implies that our 30--50\% may be overly conservative.
Conversely, it is likely that our assumption that decay photons from, for example, $\pi^{0}$'s and $\eta$'s, can be reconstructed and associated to the neutrino vertex with perfect efficiency is highly optimistic, particularly for the \faser design.
However, despite these shortcomings, by including a simple model for smearing and threshold effects, we are able to test whether the \lownu method could plausibly be useful for the FPF, and whether it is worth a more detailed future study with a full detector model and realistic simulation and reconstruction.

Using this model, as well as being able to estimate the expected event rate in the FPF flux, we are able to formulate proxy variables for \enutrue and \qz that will be detector accessible. We define the reconstructed hadronic energy, \ehadreco, as a proxy for \qz:
\begin{equation}
  \ehadreco = \Big(\sum_{i=p,\bar{p}} E^i_{\mathrm{kin}}\Big) + \Big(\sum_{i=\pi^{\pm},K^{\pm},\gamma,l^{\pm},K^{0}_{\mathrm{S}}} E^i_{\mathrm{total}}\Big),
  \label{eq:ehadreco}
\end{equation}
which is the sum of the kinetic energies of outgoing protons (and antiprotons), and the total energy of all other {\it observable particles}. Unobservable particles (neutrons, $K^0_{\mathrm{L}}$'s and neutrinos) are simply omitted. In both LE and HE GENIE simulations used throughout this study, particles that are shorter-lived than kaons are decayed by the GENIE simulation and their decay products are considered. Note that charged leptons are only included in \ehadreco if they are non-primary. E.g., the muon produced at the leptonic vertex of a $\numu$ CC interaction is not included.

The reconstructed neutrino energy, \enureco, is defined for \numu (\numub) CC events as:
\begin{equation}
  \enureco = E_{\mu} + \ehadreco,
  \label{eq:enureco}
\end{equation}
where $E_{\mu}$ is the total energy of the primary muon at the leptonic vertex.

When calculating both \enureco and \ehadreco, the energy of each particle is smeared separately according to the resolution described in Tab.~\ref{tab:detector_model}, event by event.

\begin{figure}[htbp]
  \centering
  \captionsetup[subfloat]{captionskip=-1pt}
  \subfloat[\numu--\tungsten]   {\includegraphics[width=0.9\linewidth]{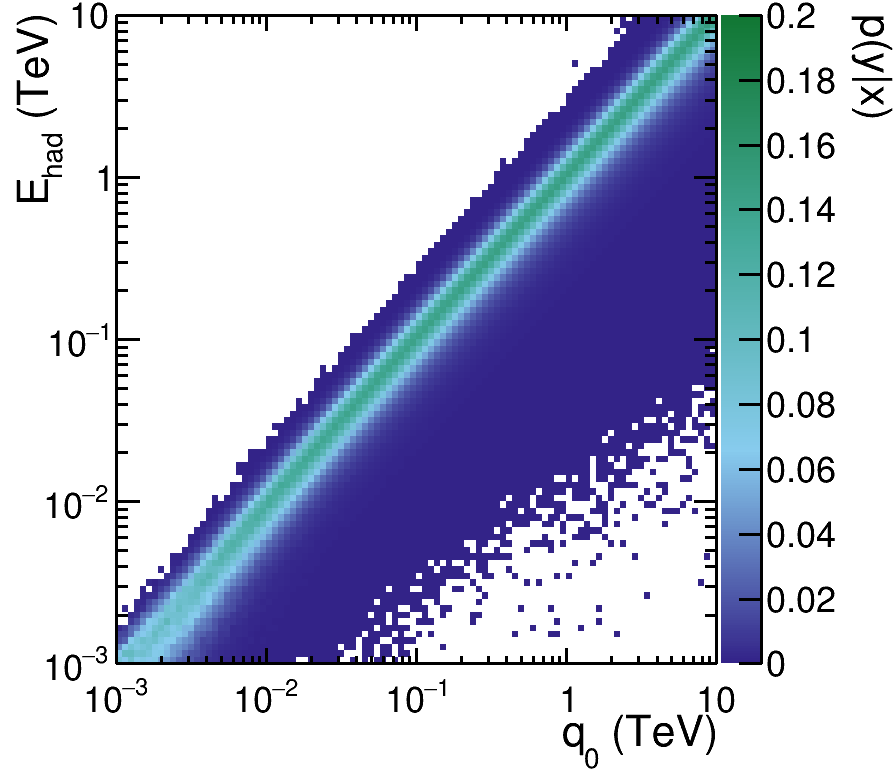}}\\\vspace{-10pt}
%  \subfloat[\numu--\argon]      {\includegraphics[width=0.45\linewidth]{figures/q0_smear_1000180400_14_10000GeV_G18_10a_02_11a.png}}\\\vspace{-10pt}
  \subfloat[\numub--\tungsten]  {\includegraphics[width=0.9\linewidth]{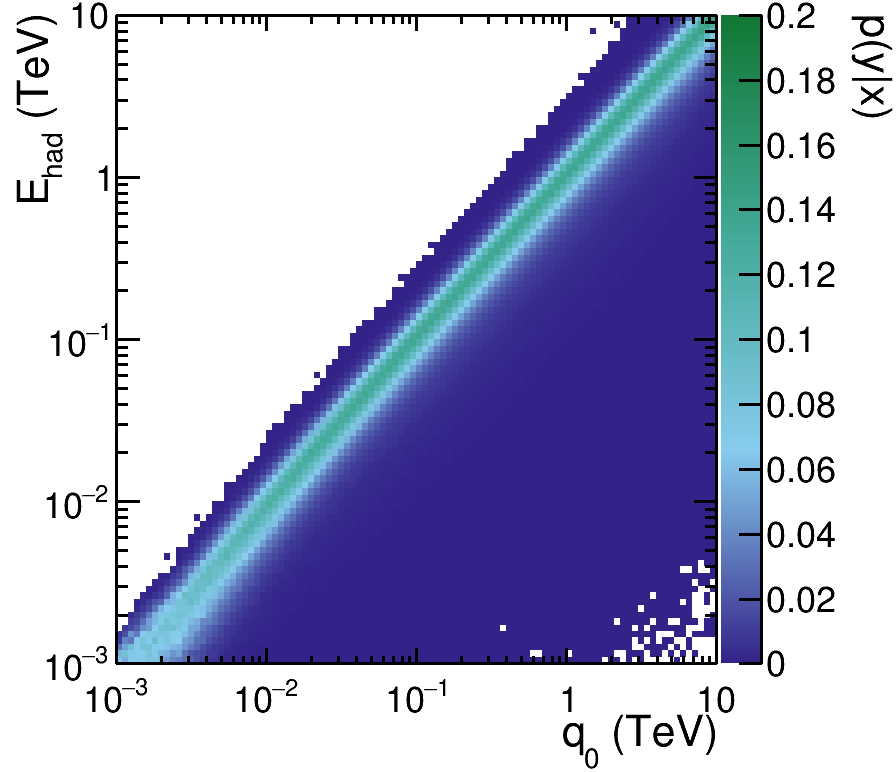}}
%  \subfloat[\numub--\argon]     {\includegraphics[width=0.45\linewidth]{figures/q0_smear_1000180400_-14_10000GeV_G18_10a_02_11a.png}}
  \caption{The \qz--\ehadreco smearing for monoenergetic 10 TeV \numu (top) and \numub (bottom) using the GENIE LE model for \faser (\tungsten). The smearing of \ehadreco uses Equation~\ref{eq:ehadreco} and the detector assumptions described in Tab.~\ref{tab:detector_model}.}
  \label{fig:ehad_smear_comp}
\end{figure}
Figure~\ref{fig:ehad_smear_comp} shows the smearing between \ehadreco and \qz for 10 TeV monoenergetic \numu and \numub CC interactions in \faser, using the detector assumptions described in Tab.~\ref{tab:detector_model} and the GENIE LE model. The bulk of both distributions shows a strong linear relationship between \qz and \ehadreco. Additionally, there is a smaller population with more pronounced smearing to lower \ehadreco, due to energy lost to unobservable particles. Although the central population is similar for \numu and \numub, the broad smearing to low \ehadreco is more significant for \numub. The smearing at 10 TeV is qualitatively similar to the smearing at other neutrino energies explored (up to kinematic limits), and for the GENIE HE model (which imposes a kinematic limit of $\qz \gtrsim 2$ GeV). The smearing is qualitatively similar between \faser (\tungsten) and FLArE (\argon), although the former has a slightly broader central $\qz \approx \ehadreco$ band than the latter, due to the larger smearing for \faser implemented in Tab.~\ref{tab:detector_model}.

\begin{figure*}[htbp]
  \centering
  \captionsetup[subfloat]{captionskip=-1pt}
  \subfloat[\numu--\tungsten]  {\includegraphics[width=0.45\linewidth]{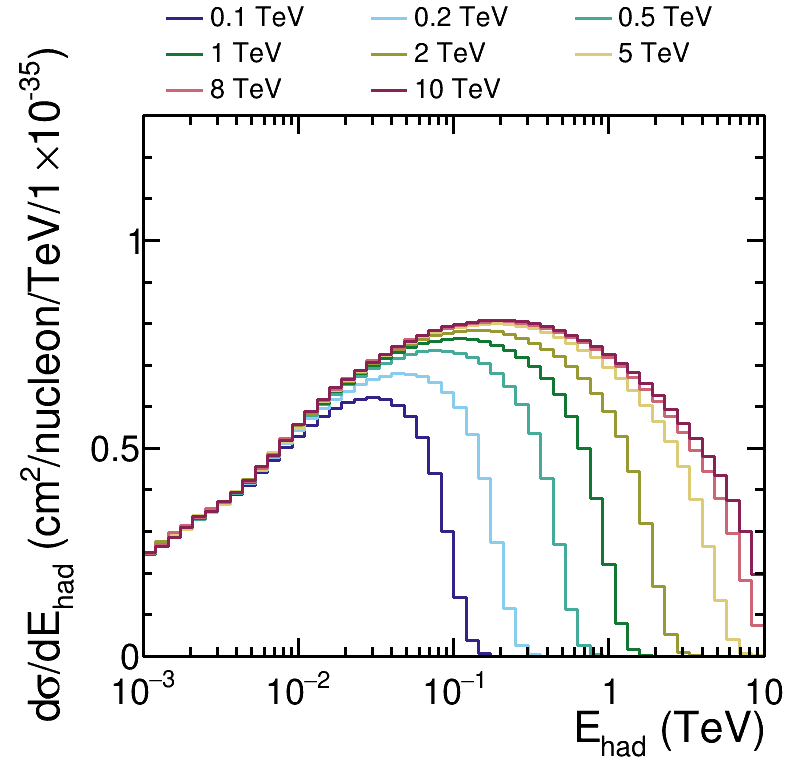}}
  \subfloat[\numu--\argon]     {\includegraphics[width=0.45\linewidth]{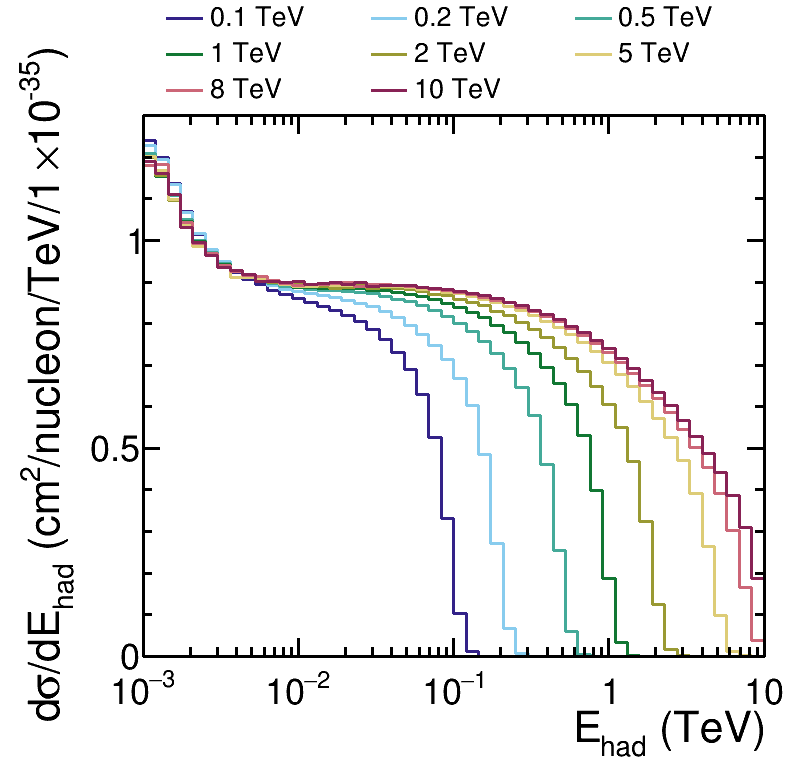}}\\\vspace{-10pt}
  \subfloat[\numub--\tungsten] {\includegraphics[width=0.45\linewidth]{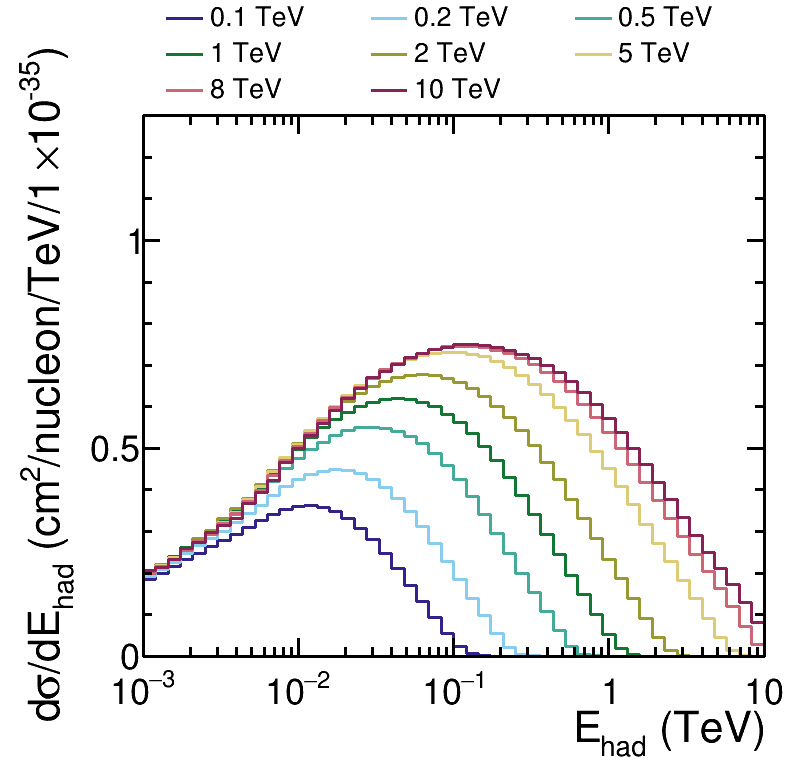}}
  \subfloat[\numub--\argon]    {\includegraphics[width=0.45\linewidth]{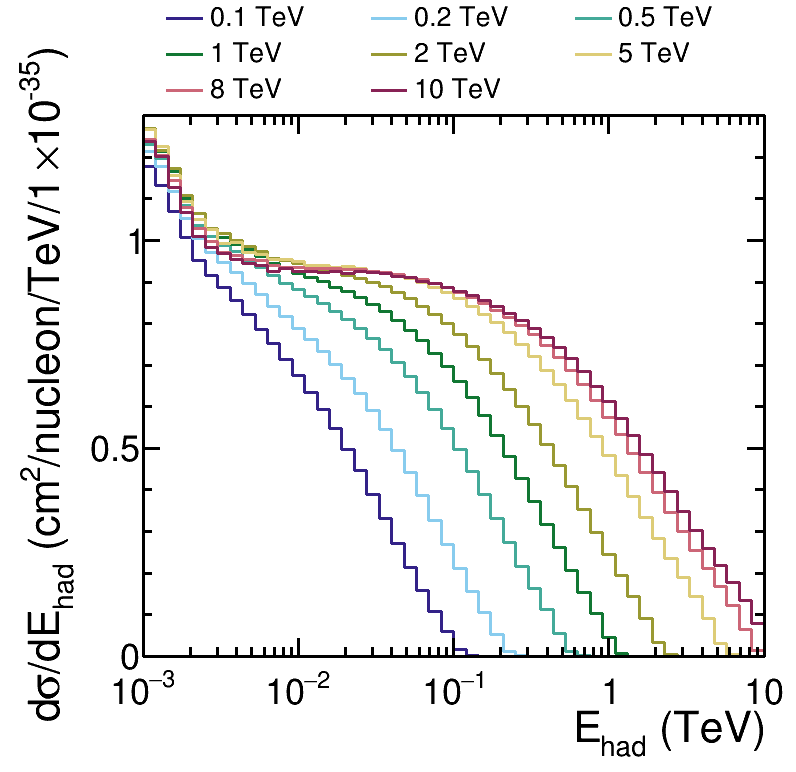}}
  \caption{The \numu (top) and \numub (bottom) CC-inclusive cross sections for both \tungsten (left) and \argon (right) targets, using the GENIE LE tune, for fixed values of \enutrue, as a function of \ehadreco.}
  \label{fig:reco_xsec_comp}
\end{figure*}
Figure~\ref{fig:reco_xsec_comp} shows the \numu (top) and \numub (bottom) CC-inclusive cross sections for both \tungsten and \argon targets, using the GENIE LE tune, for fixed values of \enutrue, as a function of \ehadreco. It can be compared to Figure~\ref{fig:true_xsec_comp}, which was shown as a function of true \qz. For both \numu--\argon and \numub--\argon, similar structures exist in both Figures~\ref{fig:reco_xsec_comp} and~\ref{fig:true_xsec_comp}, with a peak due to resonant pion production at low \ehadreco (or \qz), then a rising cross section due to DIS contributions, which tend to diverge at higher \ehadreco (or \qz) between different \enutrue histograms. There is a noticeable difference between Figures~\ref{fig:reco_xsec_comp} and~\ref{fig:true_xsec_comp} for both \numu--\tungsten and \numub--\tungsten, as the higher detector threshold and $\geq 5$ track cut suppresses the very low-\ehadreco contributions, to the extent that there is no longer a clear resonant contribution at all. 
For \numu--\argon and \numub--\argon, the resonant peak has been smeared to lower \ehadreco, compared with Figure~\ref{fig:true_xsec_comp}.
For both \numu and \numub, and both \tungsten and \argon, the cross sections appear to diverge at a lower \ehadreco for different \enutrue values in the DIS region (than the equivalent \qz values in Figure~\ref{fig:true_xsec_comp}).
In particular, both the \numub--\tungsten and \numub--\argon cross sections show $\approx$10\% differences as a function of \enutrue above the resonance region ($\ehadreco \gtrsim 3$ GeV), indicating that the \lownu method is likely to break down at lower \numub energies. Additionally, the sharp kinematic limit that was present at high-\qz in Figure~\ref{fig:true_xsec_comp}, is smeared out at high-\ehadreco in Figure~\ref{fig:reco_xsec_comp}. The general trends shown in Figure~\ref{fig:reco_xsec_comp} for the GENIE LE model are qualitatively similar to those for the HE model (with the absence of the unsimulated resonance peak at low \ehadreco).

A conclusion that may be drawn from Figure~\ref{fig:reco_xsec_comp} is that a potentially useful \lownu sample of \numu--\tungsten or \numu--\argon events at the FPF could be selected using cuts on the detector observable quantity $0 \leq \ehadreco \leq 20$ GeV, and a further restriction of $\ehadreco \geq 5$ GeV may also be placed to cut out non-DIS contributions. And whilst a potentially useful sample of \numub--\tungsten or \numub--\argon may be selected using cuts of $0 \leq \ehadreco \leq 10$ GeV, this seems less promising as energy-dependent corrections are likely to be larger. Again, for \numub interactions, a further restriction of $\ehadreco \geq 5$ GeV may also be placed to cut out non-DIS contributions.

\begin{figure*}[htbp]
  \centering
  \captionsetup[subfloat]{captionskip=-1pt}
  \subfloat[\numu--\tungsten, LE]  {\includegraphics[width=0.45\linewidth]{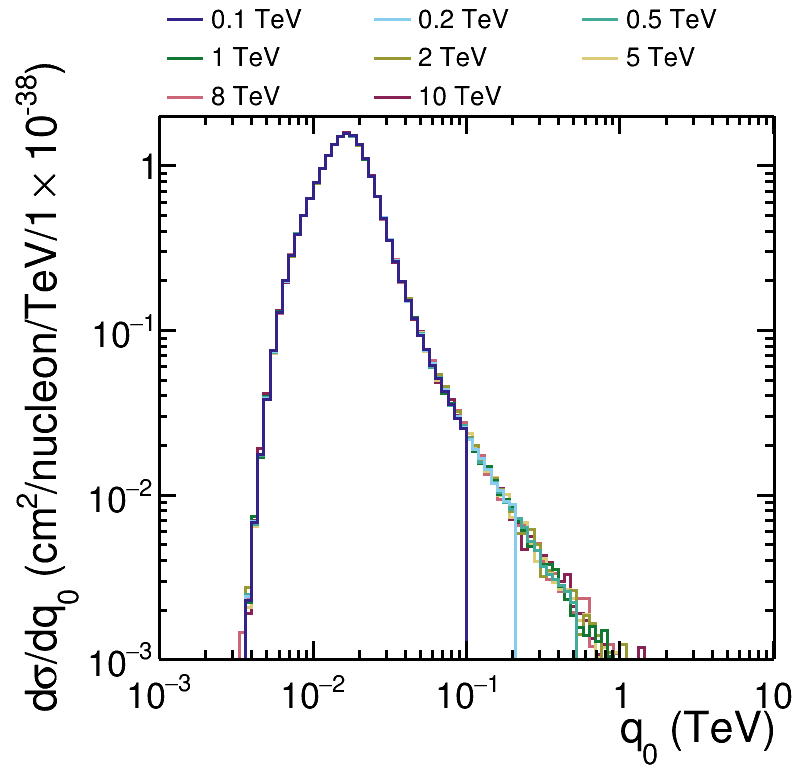}}
  \subfloat[\numu--\tungsten, HE]  {\includegraphics[width=0.45\linewidth]{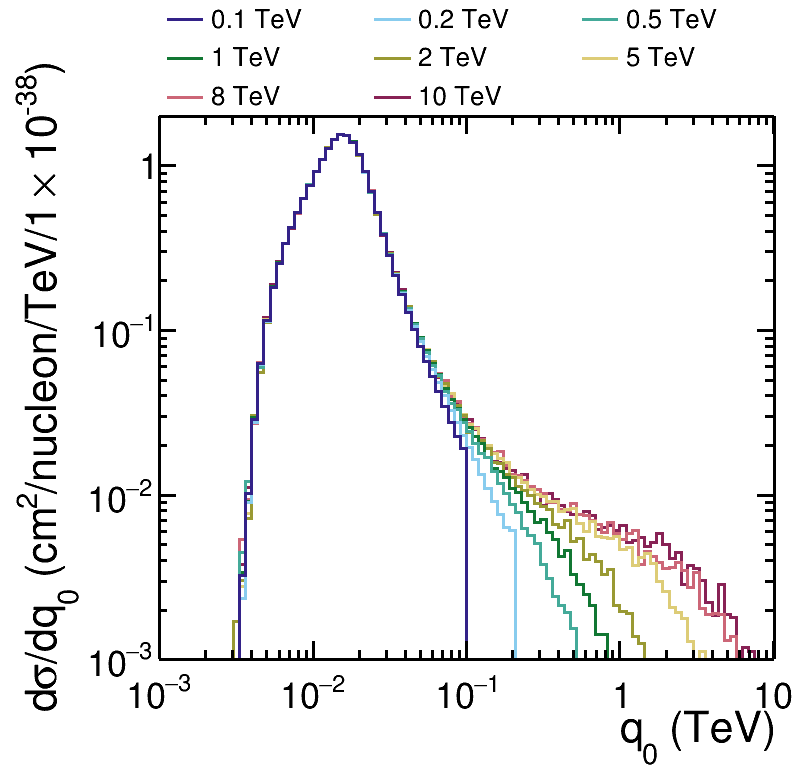}}\\\vspace{-10pt}
  \subfloat[\numub--\tungsten, LE] {\includegraphics[width=0.45\linewidth]{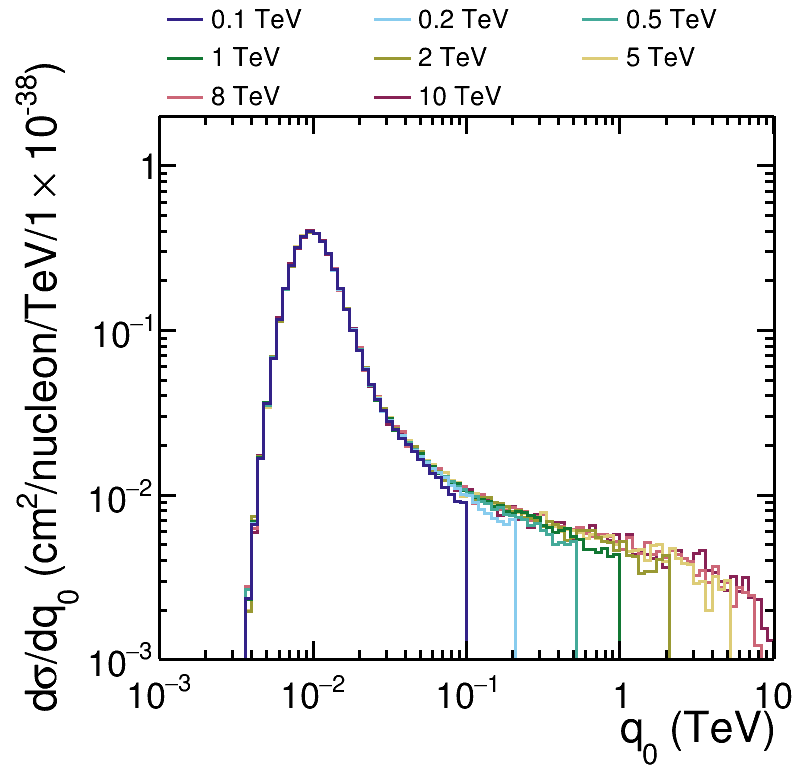}}
  \subfloat[\numub--\tungsten, HE] {\includegraphics[width=0.45\linewidth]{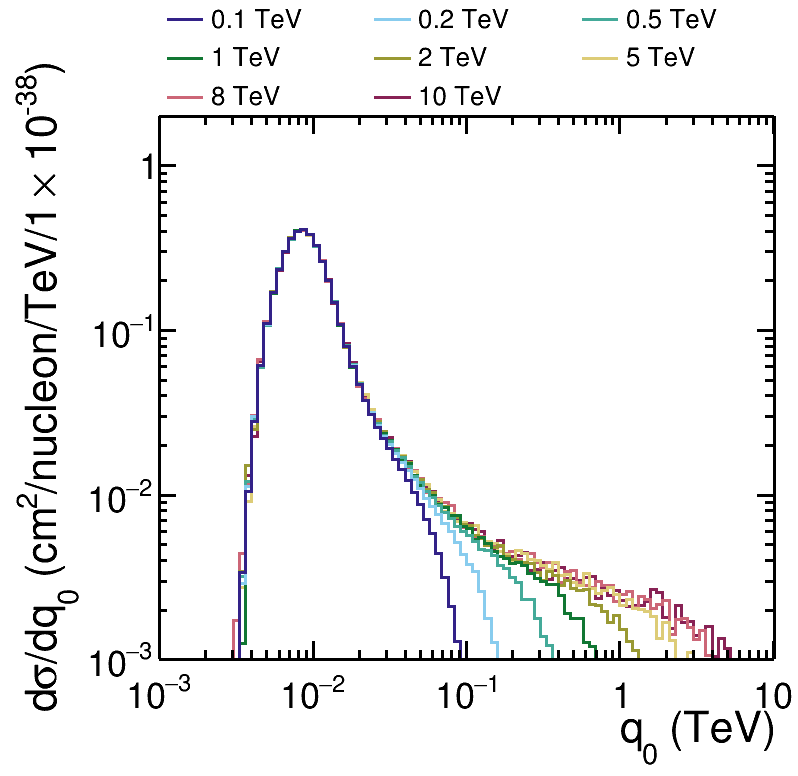}}
  \caption{Contributions to \lownu samples for both \numu--\tungsten ($5 \leq \ehadreco \leq 20$ GeV, top) and \numub--\tungsten ($5 \leq \ehadreco \leq 10$ GeV, bottom), shown as a function of true-\qz, for both the GENIE LE (left) and HE (right) models investigated in this work and for various monoenergetic values of \enutrue and normalized to a cross section.}
  \label{fig:ehad_spread_comp}
\end{figure*}
Figure~\ref{fig:ehad_spread_comp} shows the contributions to the \lownu samples defined as above, for both \numu--\tungsten ($5 \leq \ehadreco \leq 20$ GeV) and \numub--\tungsten ($5 \leq \ehadreco \leq 10$ GeV) events, as a function of true-\qz, for both LE and HE GENIE models and for different monoenergetic values of \enutrue. In all cases, the central peak has a constant cross section as a function of \enutrue. However, there is a prominent high-\qz tail for the \numub--\tungsten sample seen with both models, which contributes up to $\approx$10\% of the total cross section of the sample, increasing with \enutrue. As this tail varies with \enutrue, it will add model dependence which must be corrected if used for the \lownu method. There is a smaller tail out to high-\qz for the GENIE HE \numu--\tungsten distributions, which also varies with \enutrue, but is a few percent of the total cross section, so it adds less model dependence. The differences between the LE and HE GENIE \numu--\tungsten samples, which is also present, but less clear, for the \numub--\tungsten samples, highlights how model dependence would enter, and is largely due to the different numbers of high-\qz events for which almost all of the energy in the interaction is transferred to an unobservable particle or particles (often, but not always $K^{0}_{\mathrm{L}}$'s). The equivalent \numu--\argon and \numub--\argon samples are qualitatively similar to those shown in Figure~\ref{fig:ehad_spread_comp}, for both the GENIE LE and HE models.

\begin{figure}[htbp]
  \centering
  \captionsetup[subfloat]{captionskip=-1pt}
  \subfloat[\numu--\tungsten]  {\includegraphics[width=0.9\linewidth]{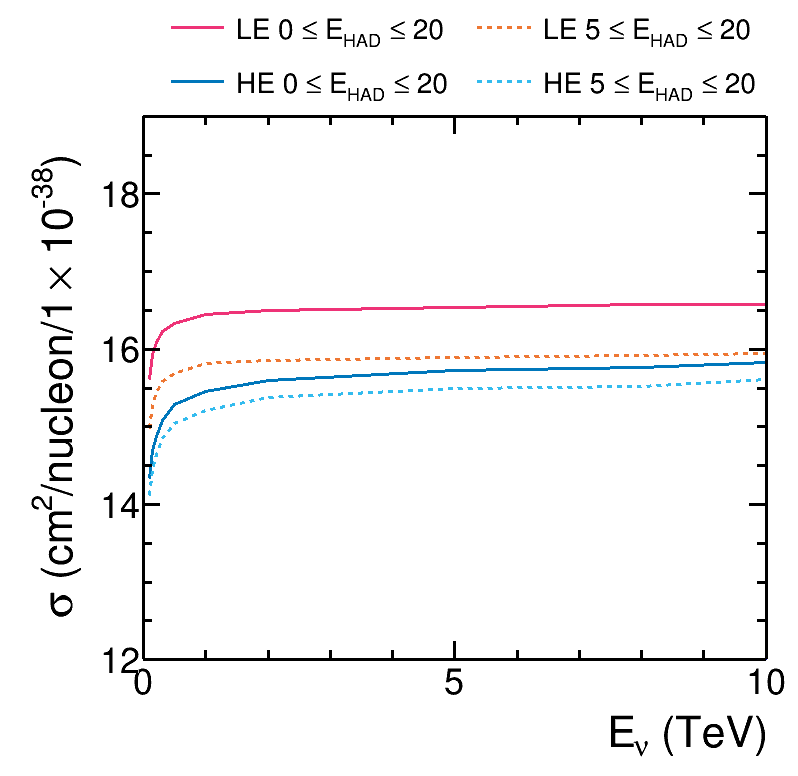}}\\\vspace{-10pt}
  %\subfloat[\numu--\argon]     {\includegraphics[width=0.45\linewidth]{figures/enu_corr_1000180400_14_ehad_xsec.png}}\\\vspace{-10pt}
  \subfloat[\numub--\tungsten] {\includegraphics[width=0.9\linewidth]{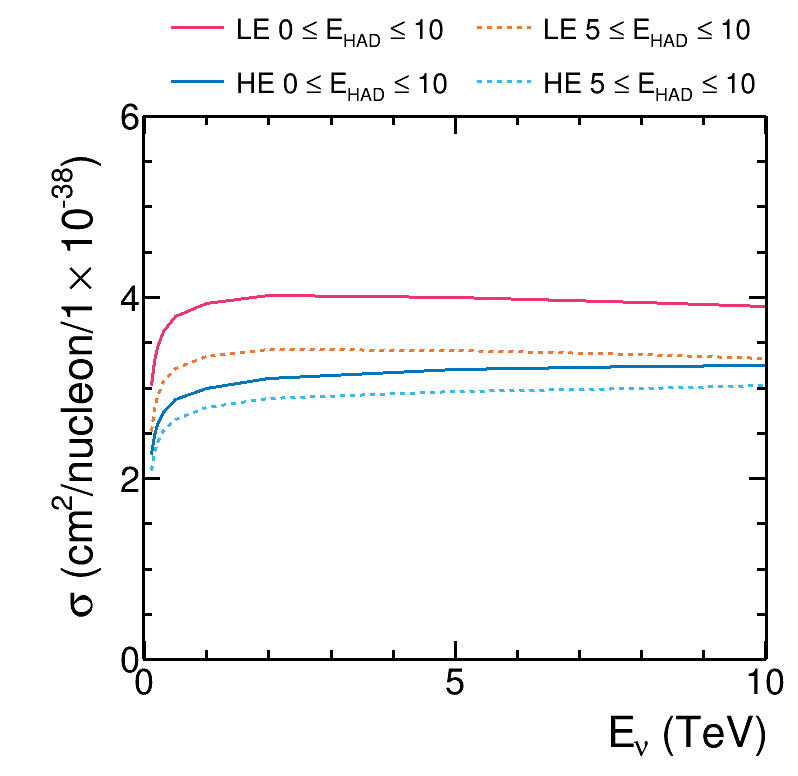}}
  %\subfloat[\numub--\argon]    {\includegraphics[width=0.45\linewidth]{figures/enu_corr_1000180400_-14_ehad_xsec.png}}  
  \caption{The \numu (top) and \numub (bottom) restricted \lownu cross section for a \tungsten target, selected using cuts on true \ehadreco, shown as a function of \enutrue for both GENIE LE and HE tunes.}
  \label{fig:lownu_ehad_xsec}
\end{figure}  
Figure~\ref{fig:lownu_ehad_xsec} shows the restricted \lownu cross section, defined here as the CC-inclusive cross section with a restriction of $\ehadreco \leq 20$ GeV ($\ehadreco \leq 10$ GeV) for \numu--\tungsten (\numub--\tungsten), as a function of \enutrue. Alternative \lownu cross sections with an additional minimum \ehadreco restriction of $\ehadreco \geq 5$ GeV are included. The general behavior for \numu--\argon and \numub--\argon is very similar. These should be compared to the \lownu cross sections obtained with equivalent cuts on true-\qz in Figure~\ref{fig:lownu_q0_xsec}. For both \tungsten and \argon, the \ehadreco selected \lownu sample cross sections are $\approx$10\% larger than their \qz selected counterparts for \numu, but up to $\approx$50\% larger for \numub, consistent with the large migration from high-\qz into the sample shown in Figure~\ref{fig:ehad_spread_comp}.

For \numu--\tungsten, the cross sections in Figure~\ref{fig:lownu_ehad_xsec} are relatively flat, with corrections of up to $\approx$5\% for the LE model and $\approx$10\% for the HE model, as a function of neutrino energy. Similar behavior is seen for \numu--\argon. The model difference of $\approx$5\% gives a sense of the size of the model dependence of these corrections, and is larger than the subpercent differences seen with samples selected with cuts on true \qz. The cross section is significantly less flat for \numub--\tungsten (and \numub--\argon), with \enutrue-dependent differences of up to $\approx$30\% for the HE model, and $\approx$15\% for the LE model. There are large, 10--15\% differences between the two models across much of the energy range of interest, implying significant model dependence for the application of the \lownu method for the \numub FPF flux. The additional restriction of $\ehadreco \geq 5$ GeV does not significantly change any of these conclusions.

Figure~\ref{fig:lownu_ehad_xsec} demonstrates that the second requirement for the \lownu method to work, as identified in Section~\ref{sec:lownu}, is fulfilled for \numu interactions --- a \lownu appropriate sample can be selected in FPF detectors without introducing significant model dependence. The conclusion for \numub interactions is less robust, as it seems likely that some significant model-dependence will be introduced.

\begin{figure}[htbp]
  \centering
  \captionsetup[subfloat]{captionskip=-1pt}
  \subfloat[\faser]  {\includegraphics[width=0.9\linewidth]{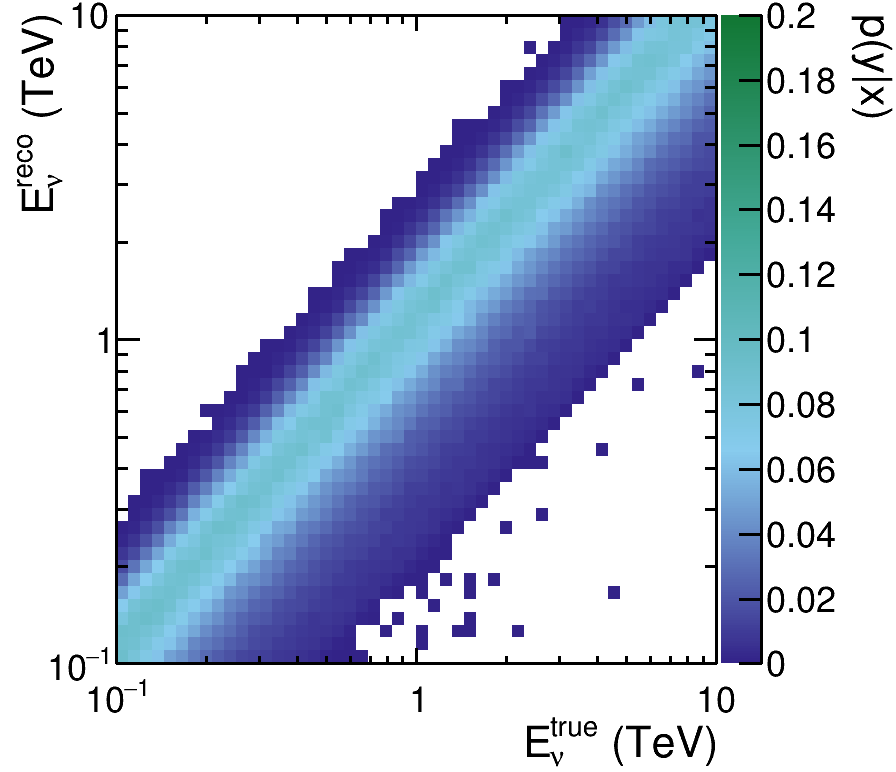}}\\\vspace{-10pt}
  \subfloat[FLArE]   {\includegraphics[width=0.9\linewidth]{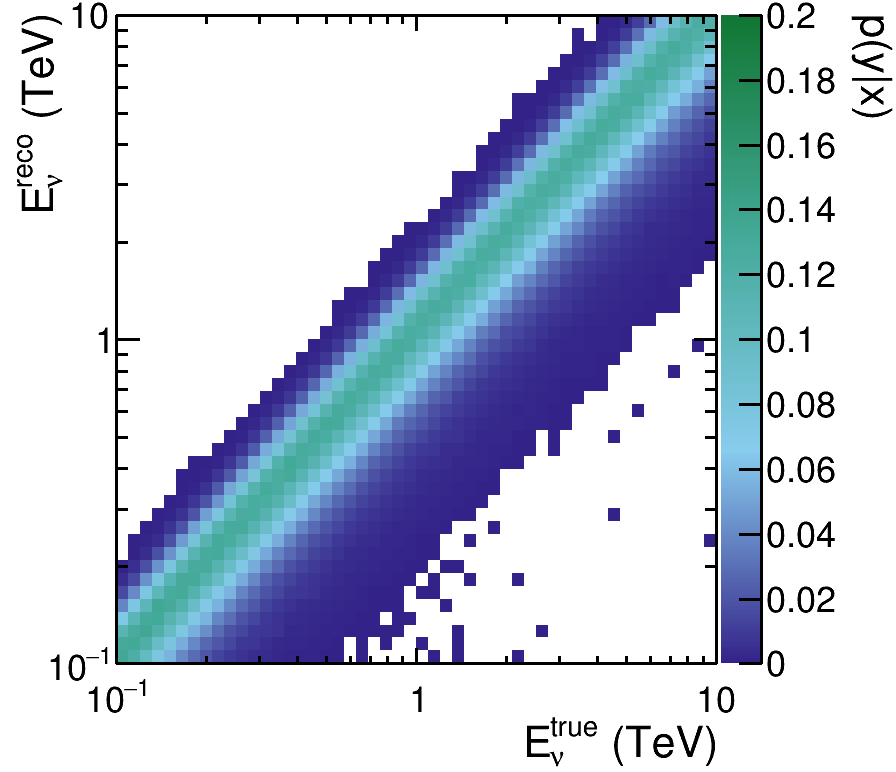}}
  \caption{\enutrue--\enureco smearing using the detector assumptions described for \faser (\tungsten) and FLArE (\argon) in Tab.~\ref{tab:detector_model}, shown for \numu and the GENIE LE tune. The smearing is similar for both GENIE models and for both \numu and \numub.}
  \label{fig:enu_smear_comp}
\end{figure}
Figure~\ref{fig:enu_smear_comp} shows the smearing between \enutrue and \ehadreco for both \faser (\tungsten) and \flare (\argon) for \numu CC \lownu interactions selected using cuts of $5 \leq \ehadreco \leq 20$ GeV, using the detector assumptions described in Tab.~\ref{tab:detector_model} and the GENIE LE model. The samples used to produce these plots had a $\sim$1/\enutrue flux in the range $0.1 \leq \enutrue \leq 10$ TeV, which produces approximately equal statistics in the \lownu sample across the energy range. There is a strong linear relationship between \enutrue and \enureco for both detector models, although there is broad smearing, which is more pronounced for \faser than FLArE, due to the larger smearing and higher thresholds for \faser implemented in Tab.~\ref{tab:detector_model}. Qualitatively similar smearing was observed for \numub samples for both detector models (for a \lownu sample selected using cuts of $5 \leq \ehadreco \leq 10$ GeV).

\begin{figure}[htbp]
  \centering
  \captionsetup[subfloat]{captionskip=-1pt}
  \subfloat[\faser]   {\includegraphics[width=0.8\linewidth]{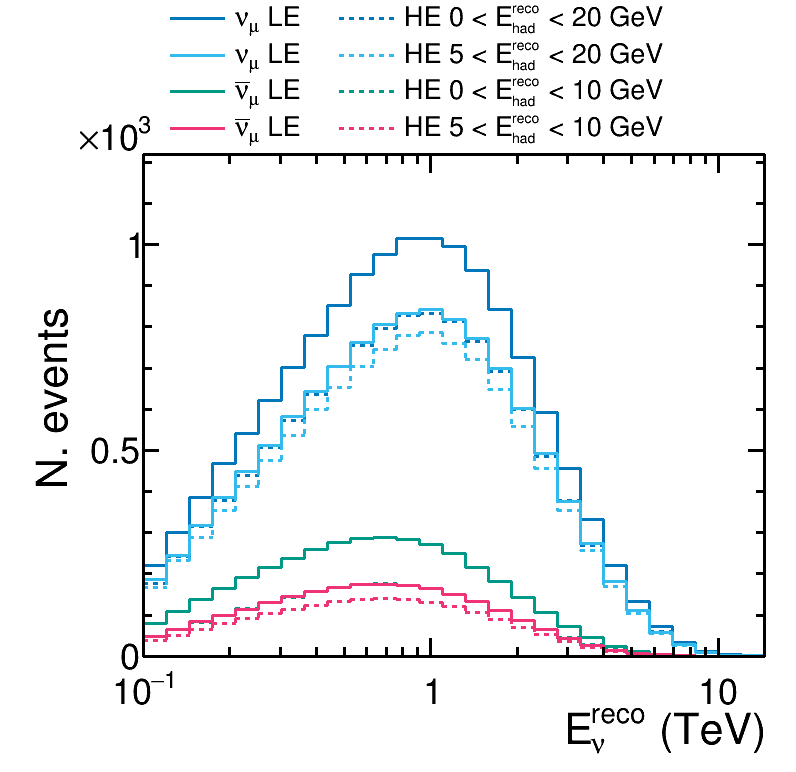}}\\\vspace{-10pt}
  \subfloat[FLArE10]  {\includegraphics[width=0.8\linewidth]{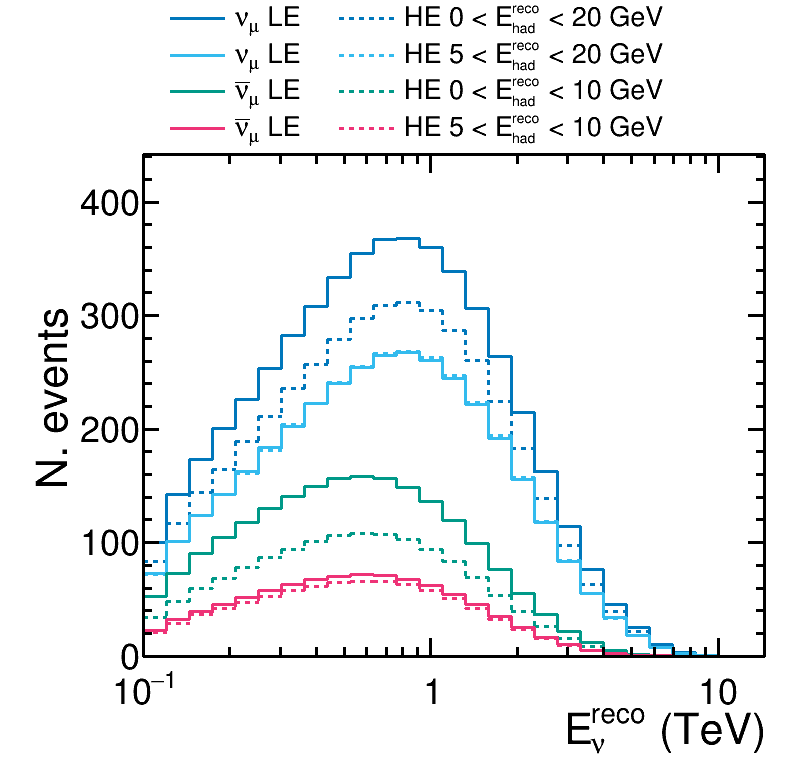}}\\\vspace{-10pt}
  \subfloat[FLArE100] {\includegraphics[width=0.8\linewidth]{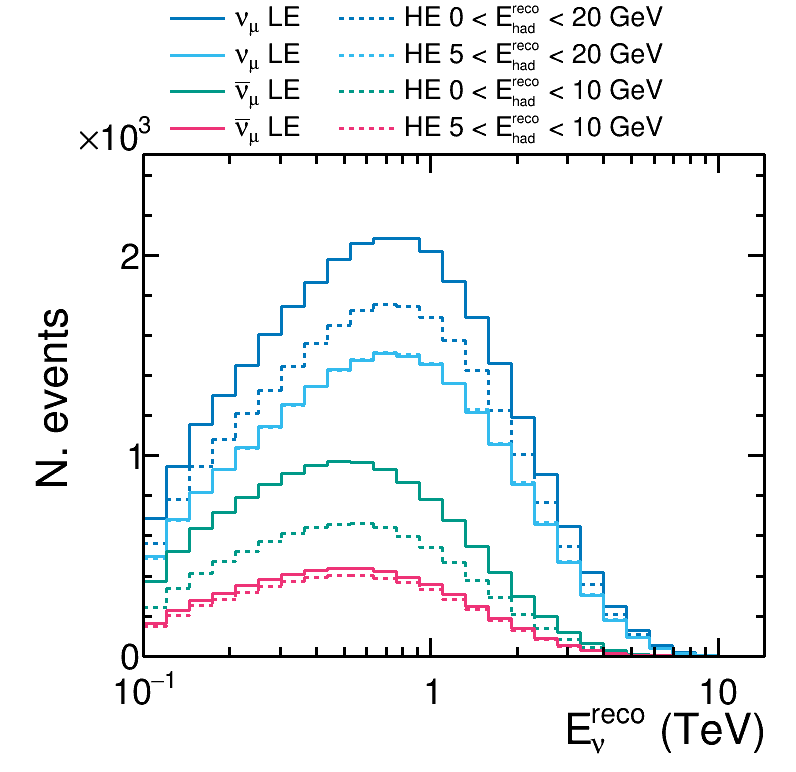}}
  \caption{Candidate CC \lownu sample event rates for \numu scattering in \faser, FLArE10 and FLArE100 for their full 3000 fb$^{-1}$ anticipated runs, as a function of \enureco, shown for two alternative \lownu sample selection cuts, $0 \leq \ehadreco \leq 20$ GeV and $5 \leq \ehadreco \leq 20$, for both LE and HE GENIE tunes, produced using the fluxes provided by Ref.~\cite{Kling:2021gos}.}
  \label{fig:reco_rate_comp}
\end{figure}
Figure~\ref{fig:reco_rate_comp} shows the expected \numu and \numub \lownu sample event rates for the three FPF detector options considered here, \faser, FLArE10 and FLArE100 for their full 3000 fb$^{-1}$ anticipated runs, as a function of \enureco, using the nominal flux predictions from Ref.~\cite{Kling:2021gos}. Two possible \numu (\numub) \lownu samples are considered for completeness, one with a restriction of $\ehadreco \leq 20$ GeV ($\ehadreco \leq 10$ GeV) and the other with $5 \leq \ehadreco \leq 20$ GeV ($5 \leq \ehadreco \leq 10$ GeV), the latter of which removes non-DIS contributions. Both the LE and HE GENIE tunes described in Section~\ref{sec:model} are shown, and the full detector model described in Tab.~\ref{tab:detector_model} is applied. All three detector options provide $\mathcal{O}$(10,000) \numu events and $\mathcal{O}$(1,000) \numub events, for all models and candidate cut values. The \lownu samples shown in Figure~\ref{fig:reco_rate_comp} correspond to $\approx$1\% of the total CC-inclusive event rates (shown in Figure~\ref{fig:total_rate_comp}).

Figure~\ref{fig:reco_rate_comp} demonstrates that the third requirement for the \lownu method to work, as identified in Section~\ref{sec:lownu}, is fulfilled. The FPF detectors have to produce a large enough sample of \lownu events, selected experimentally, to constrain the flux --- at least the \numu flux.

% \FloatBarrier
\section{Example fit for extracting the neutrino flux shape}
\label{sec:fit}

\begin{figure*}[htbp]
  \centering
  \captionsetup[subfloat]{captionskip=-1pt}
  \includegraphics[width=0.8\linewidth]{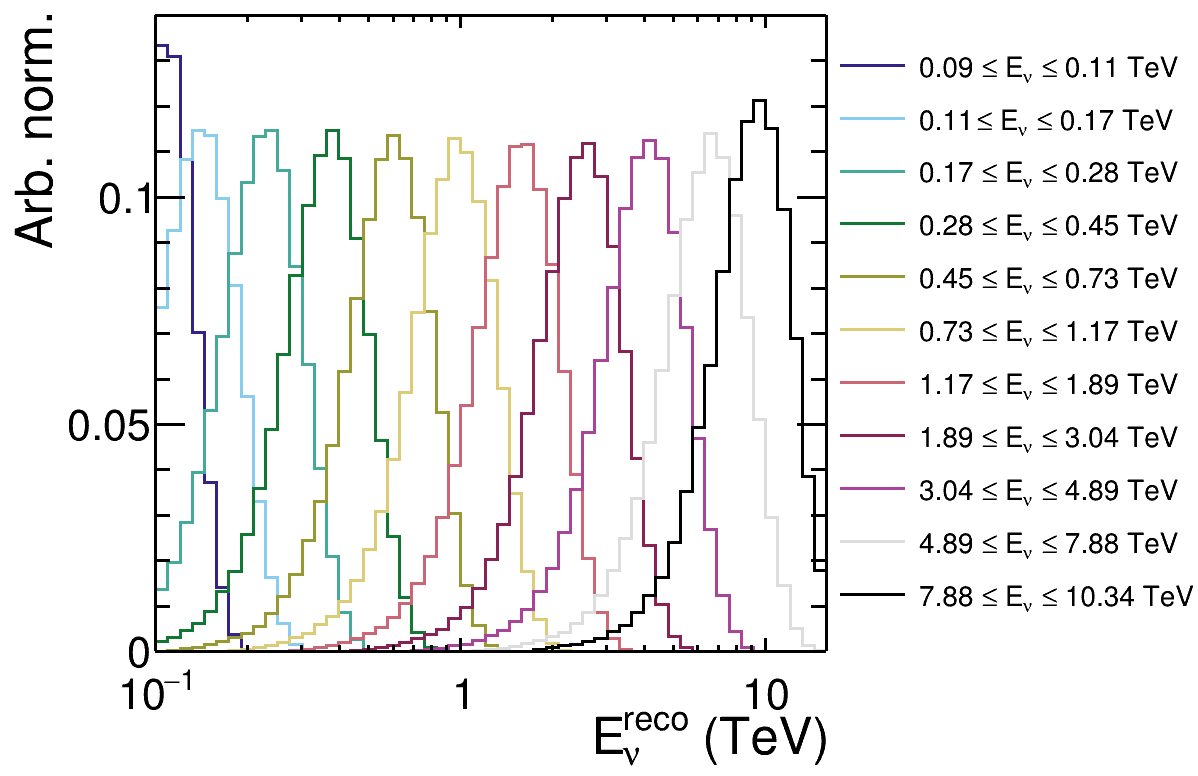}
  \caption{\enutrue templates for the FLArE (10 or 100) \numu--\argon CC $5 \leq \ehadreco \leq 20$ \lownu sample, shown as a function of \enureco. Each template represents the predicted \enureco distribution for the indicated \lownu sample, with \enutrue flat in the region indicated by the legend.}
  \label{fig:fit_templates}
\end{figure*}
In order to quantify the power of the \lownu method for the FPF, here we extract a shape-only flux constraint on the \numu flux using the assumptions described in Section~\ref{sec:reco}, and a template fit approach. Detailed detector systematic uncertainties have not been included, and would need to be evaluated once the FPF detector design is finalized. However, this analysis demonstrates the power of the method and could be extended to incorporate systematic uncertainties in the future.

The principle of the template likelihood fit used here is simple. Two independent Monte Carlo (MC) samples are generated, one which follows the SIBYLL v2.3 flux for the given FPF detector~\cite{Kling:2021gos}, and one which is independent of the flux shape. For the nominal MC sample, the \lownu selection cuts (described in Section~\ref{sec:reco}) are applied and binned as a function of \enureco, as can be seen in Figure~\ref{fig:reco_rate_comp}. This corresponds to the expected \lownu sample in the relevant detector, assuming that the nominal flux is correct. The independent sample is divided into ``templates'' which each correspond to an \enutrue range, and each template is binned as a function of \enureco. Varying the normalization of each template (which can be thought of as a flux bin in \enutrue), and summing their contributions gives a predicted \enureco spectrum for a given flux. The predicted and nominal \enureco spectra can then be compared through a negative log-likelihood test statistic:
\begin{equation}
  \chi^{2} = 2\sum^{N}_{i=1}\left[ \mu_{i}(\vec{\mathbf{x}}) - n_{i} + n_{i}\ln\frac{n_{i}}{\mu_{i}(\vec{\mathbf{x}})}\right],
  \label{eq:chi2}
\end{equation}
\noindent where $i$ indicates the \enureco bin (of which there are $N$), $\mu_{i}(\vec{\mathbf{x}})$ is the independent MC prediction, which is a function of the template normalizations, $\vec{\mathbf{x}}$, and $n_{i}$ is the number of events in the nominal simulation (or data in a real analysis). By minimizing the value of the $\chi^{2}$ test statistic given in Equation~\ref{eq:chi2}, with respect to $\vec{\mathbf{x}}$, a best fit flux distribution in \enutrue can be extracted. For this analysis, the minimization of Equation~\ref{eq:chi2} was carried out using the MIGRAD algorithm in the ROOT implementation~\cite{root} of the MINUIT package~\cite{minuit}.

Example templates are shown for the FLArE (10 or 100) \numu--\argon CC $5 \leq \ehadreco \leq 20$ GeV \lownu sample in Figure~\ref{fig:fit_templates}. All templates were produced using a large ($\mathcal{O}$(100) million event) GENIE samples for each detector, neutrino flavor and GENIE model configuration, using a flux that fell as 1/\enutrue. This falling flux was used to ensure that the \lownu samples had approximately equal MC statistics as a function of \enutrue (a flat flux would not have), although this factor was then removed by applying a weight for each event in the analysis. The resulting templates have minimal MC statistical uncertainties across their entire range. The \enutrue boundaries used was motivated by the binning in which the FPF flux is provided in (see Ref.~\cite{Kling:2021gos}), although it is worth stressing that this is for presentational purposes, not because the analysis requires any particular binning.

\begin{figure}[htbp]
  \centering
  \captionsetup[subfloat]{captionskip=-1pt}
  \includegraphics[width=0.9\linewidth]{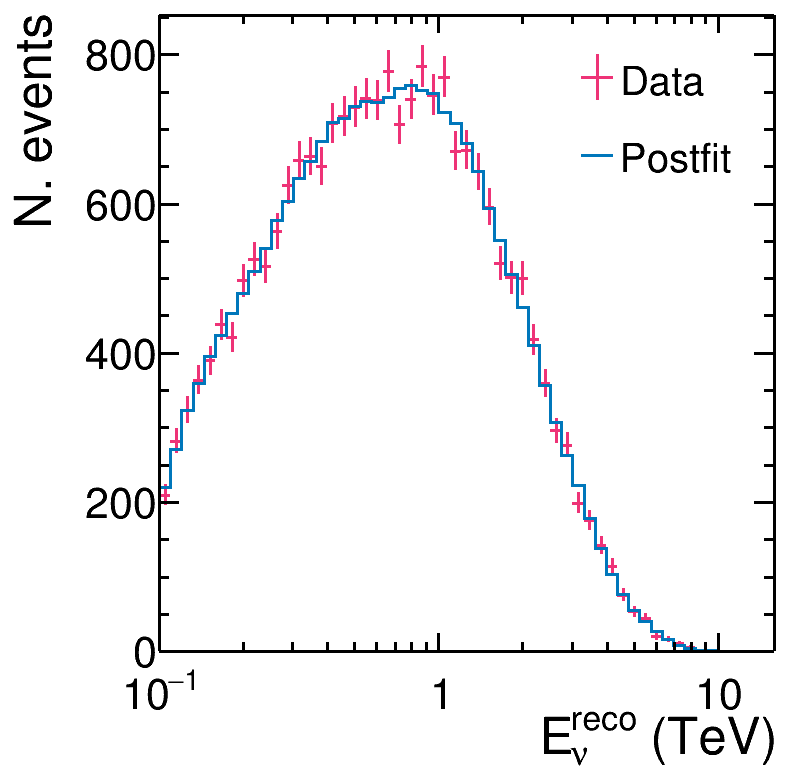}
  \caption{Example template likelihood fit result for the FLArE100 detector model and flux. The ``data'' is a prediction for the CC $5 \leq \ehadreco \leq 20$ GeV \lownu sample made using the FLArE100 flux from Ref.~\cite{Kling:2021gos}, corresponding to 3000 fb$^{-1}$, a statistical throw has been included. The postfit shows the best fit \enureco distribution produced by minimizing the test statistic described in Equation~\ref{eq:chi2} with respect to the template normalizations, using the tempate distributions shown in Figure~\ref{fig:fit_templates}.}
  \label{fig:fit_bestfit}
\end{figure}
An example result from a single template fit to the FLArE100 \numu CC $5 \leq \ehadreco \leq 20$ GeV \lownu sample, binned as a function of \enureco, is shown in Figure~\ref{fig:fit_bestfit}, corresponding to an integrated luminosity of 3000 fb$^{-1}$. The ``data'' was generated according to the FLArE100 flux provided by Ref.~\cite{Kling:2021gos} (as shown in Figure~\ref{fig:flux_comp}), and is independent of the MC generated to make the templates as described. The MC statistical uncertainties for this ``data'' sample are negligible, but a statistical throw has been included by drawing from a Poisson distribution around the nominal prediction, in order to mimic real data. As expected, the postfit result, which is obtained by varying the normalization of the templates shown in Figure~\ref{fig:fit_templates} as described by Equation~\ref{eq:chi2}, agrees well with the data. The highest energy template, $7.88 \leq \enutrue \leq 10.34$ TeV was excluded from this, and all other fits, because the corresponding best-fit template normalization was consistently 0, which is problematic as it is a fit boundary. Given the extremely small (although nonzero) flux and event rate in this true energy bin (see Figures~\ref{fig:flux_comp} and~\ref{fig:total_rate_comp}), this does not affect the general interpretation of these results.

When fit to the \lownu sample, the postfit template normalizations correspond to the shape of the neutrino energy spectrum as a function of true neutrino energy, with the caveat that any energy-dependent corrections to the cross section of the \lownu sample (see Figure~\ref{fig:lownu_ehad_xsec}) must also be applied. The statistical uncertainty due to the expected data statistics can be extracted by carrying out an ensemble of fits, where each fit in the ensemble has an independent statistical throw of the data.

\begin{figure}[htbp]
  \centering
  \captionsetup[subfloat]{captionskip=-1pt}
  \subfloat[FLArE100 \numu]  {\includegraphics[width=0.9\linewidth]{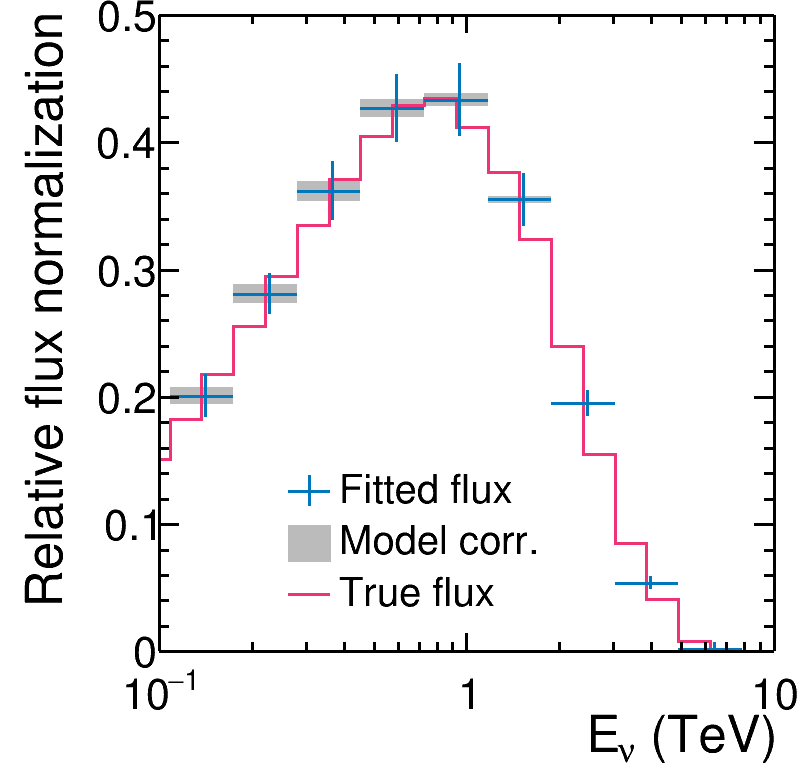}}\\\vspace{-10pt}
  \subfloat[FLArE100 \numub] {\includegraphics[width=0.9\linewidth]{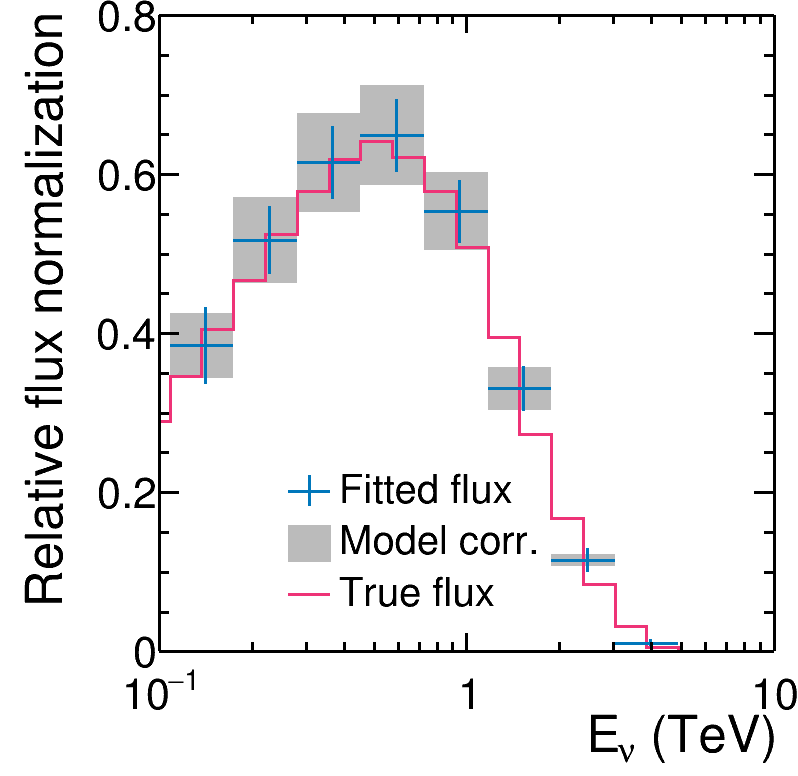}}
  \caption{Best fit values of the fitted flux distributions obtained from fits to the expected FLArE100 \numu (\numub) CC $5 \leq \ehadreco \leq 20$ GeV ($5 \leq \ehadreco \leq 10$ GeV) \lownu samples. The diagonals of the shape-only postfit covariance matrix are shown on the fitted flux. The model correction uncertainty shows the difference between using the LE and HE models to correct for \enutrue-dependent effects. The true flux is the nominal flux used to generate the ``data'' distributions, taken from Ref.~\cite{Kling:2021gos} (see Figure~\ref{fig:flux_comp}).}
  \label{fig:fit_extracted_flux}
\end{figure}
\begin{figure}[htbp]
  \centering
  \captionsetup[subfloat]{captionskip=-1pt}
  \includegraphics[width=0.9\linewidth]{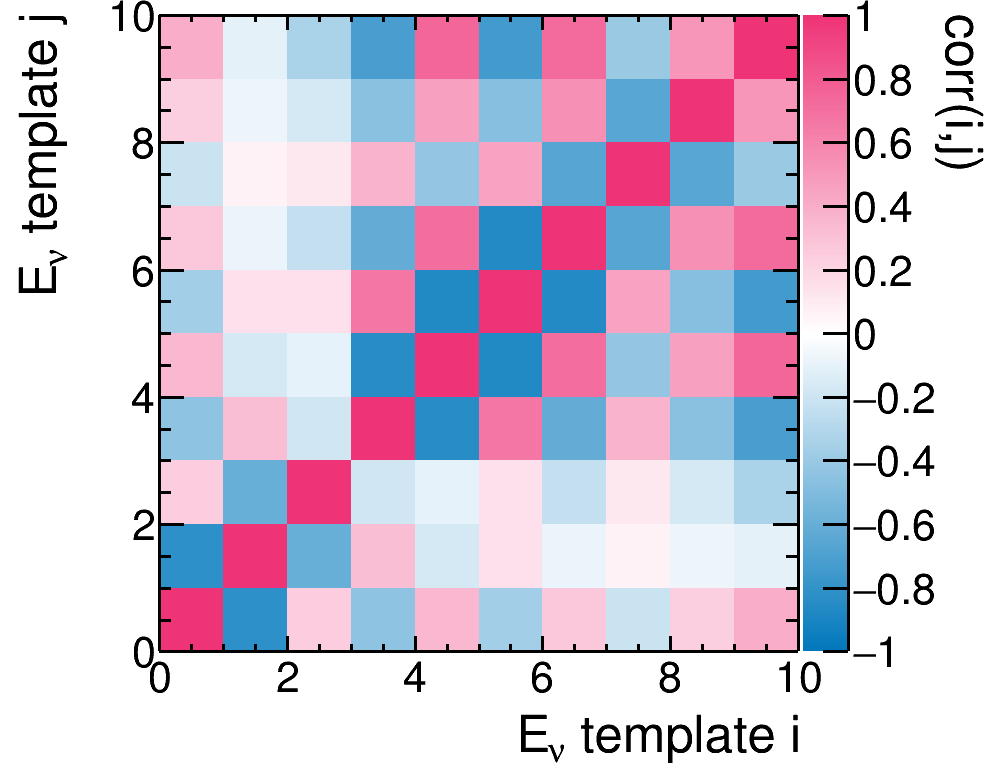}
  \caption{The shape-only correlation matrix obtained from an ensemble of 5000 template likelihood fits to the FLArE100 sample, where each fit has a statistically independent throw of the ``data'' distribution.}
  \label{fig:fit_shape_covar}
\end{figure}
Figure~\ref{fig:fit_extracted_flux} shows the value of the postfit template parameters with energy-dependent corrections applied, for both FLArE100 \numu and \numub samples, with an ensemble of 5000 fits in which a separate statistical throw was carried out for the data in each throw. The true \numu and \numub fluxes are also shown (magenta lines), and the postfit flux obtained with the \lownu fits and the true flux shapes agree well in both cases. The uncertainties on the blue points correspond to the diagonals of the shape-only covariance matrix extracted from the ensemble of 5000 fits. The shape-only {\it correlation} matrix for the ensemble of FLArE100 \numu fits is shown in Figure~\ref{fig:fit_shape_covar}. The extracted value of template 0 was included in the fits, and is shown in Figure~\ref{fig:fit_shape_covar}, but is not shown in Figure~\ref{fig:fit_extracted_flux} as the energy-dependent correction factor is large and uncertain at the lowest energies.

Figure~\ref{fig:fit_extracted_flux} also shows the uncertainty on the extracted flux prediction due to the choice of GENIE model used to apply the energy-dependent correction, which is taken as the full difference between using the LE and HE corrections bin-by-bin. For the \numu flux extraction, this model-dependent uncertainty is small, considerably smaller than the statistical uncertainty on the extracted flux shape. This is not the case for the \numub flux extraction, in which the model-dependent uncertainty is larger than the statistical uncertainty across most of the bins. In both cases, the uncertainty reduces with increasing \enutrue (as can be understood from Figure~\ref{fig:lownu_ehad_xsec}). Although the two models investigated in this work may not fully span the model-dependent uncertainty, the small potential for bias we observe for \numu suggests that the \lownu method will be an important tool for constraining the \numu flux-shape at the FPF. However, this is less likely to be the case for the \numub flux shape, although it may still be a valuable cross check, given the extremely large flux uncertainties expected at the FPF, particularly in the high-\enutrue region where charmed hadron decays dominate.

\begin{figure}[htbp]
  \centering
  \captionsetup[subfloat]{captionskip=-10pt}
  \subfloat[\faser]   {\includegraphics[width=0.8\linewidth]{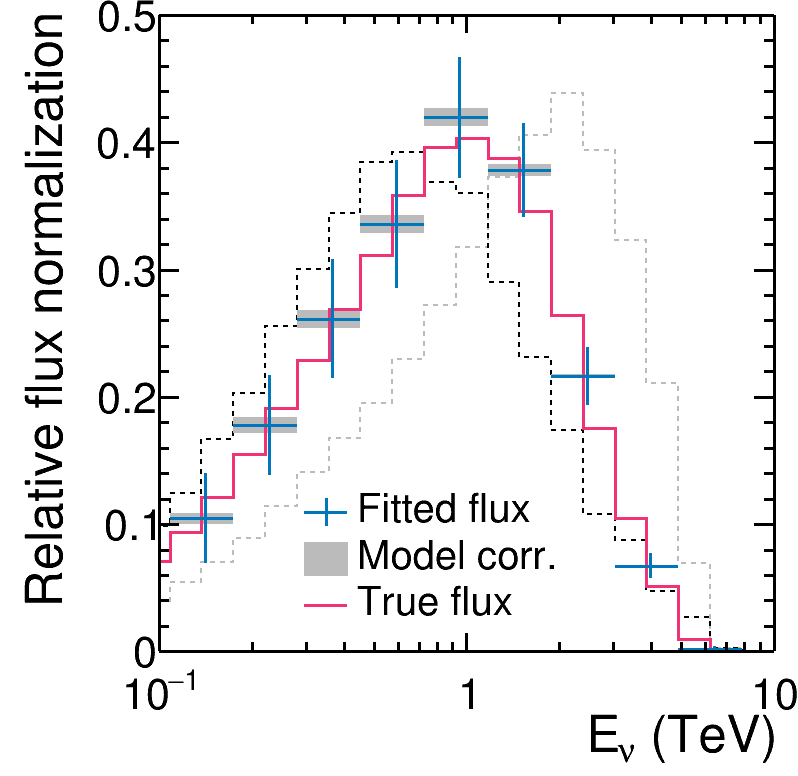}}\\\vspace{-10pt}
  \subfloat[FLArE10]  {\includegraphics[width=0.8\linewidth]{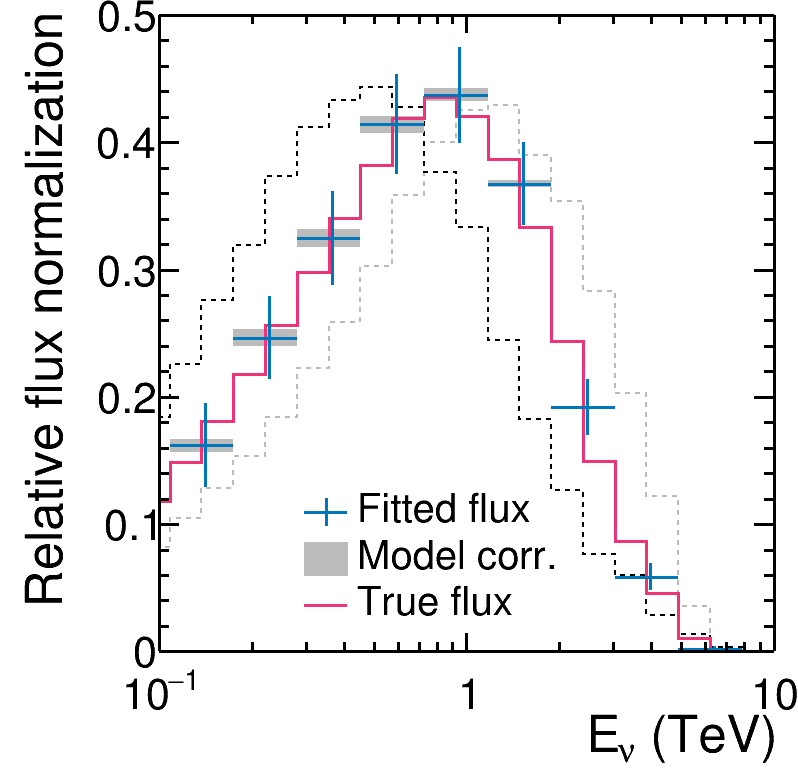}}\\\vspace{-10pt}
  \subfloat[FLArE100] {\includegraphics[width=0.8\linewidth]{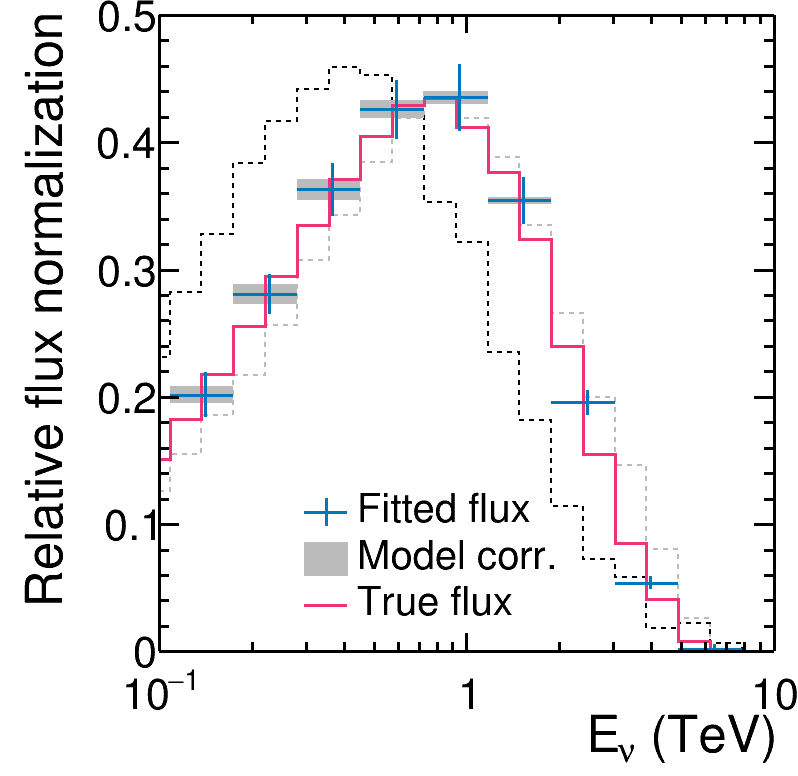}}\vspace{-5pt}
  \caption{Best-fit values of the fitted flux distributions obtained from fits to the \faser, FLArE10 and FLArE100 \numu CC $5 \leq \ehadreco \leq 20$ GeV \lownu samples. The diagonals of the shape-only postfit covariance matrix are shown on the fitted flux. The model correction uncertainty shows the difference between using the LE and HE models to correct for \enutrue-dependent effects. The true flux is the nominal flux used to generate the ``data'' distributions, taken from Ref.~\cite{Kling:2021gos} (see Figure~\ref{fig:flux_comp}), which uses the SIBYLL v2.3d model for both light and charmed hadron production. The dashed black (gray) line shows an alternative flux which uses the EPOSLHC (DPMJET-3.2019.1) model for light (charmed) hadron production.}
  \label{fig:postfit_flux_comp}
\end{figure}
Figure~\ref{fig:postfit_flux_comp} shows the extracted flux shapes obtained through an ensemble of 5000 template likelihood fits to the \faser, FLArE10 and FLArE100 \numu CC $5 \leq \ehadreco \leq 20$ GeV \lownu samples. In each case, there is good agreement between the postfit flux shape (blue histograms) and the true flux (magenta lines). The diagonals of the shape-only postfit covariance matrix are shown on the fitted flux. The model correction uncertainty shows the difference between using the LE and HE models to correct for \enutrue-dependent effects. Although the fluxes sampled by each detector differ, as do their detector masses, all three options considered here produce 5--10\% bin-to-bin relative flux uncertainties, considerably smaller than the a priori flux shape uncertainty from simulations of neutrino production in the forward region~\cite{Kling:2021gos}. Additionally, as the bins are strongly correlated (see the example for FLArE100 in Figure~\ref{fig:fit_shape_covar}), the diagonals of the postfit covariance as shown in Figure~\ref{fig:postfit_flux_comp} do not fully reflect the ability to discriminate between different flux shapes provided by this method.

Figure~\ref{fig:postfit_flux_comp} also shows the flux shapes for two alternative flux models, the EPOSLHC model~\cite{Pierog:2013ria} for light hadron production with the SIBYLL v2.3d model~\cite{Riehn:2019jet} for charmed hadron production (black dashed line), and the DPMJET-3.2019.1 model~\cite{Roesler:2000he} for charmed hadron production with the SIBYLL light hadron production (gray dashed line). These alternative models produce very different flux shapes to the nominal model (magenta solid line) which uses SIBYLL for both light and charmed hadron production. All of the fluxes used here are provided in Ref.~\cite{Kling:2021gos}. It is clear that for all of the detector options considered here and the integrated luminosity of 3000 fb$^{-1}$, the constraint on the flux shape from the \lownu samples described above would be able to differentiate between these three flux models.

\section{Conclusions}
\label{sec:conclusions}
The neutrino flux expected at the FPF and the neutrino cross section at FPF energies are both unknown. This presents an exciting opportunity, as there is a broad range of measurements which the FPF will be able to make in the neutrino sector. But also a significant challenge, as the FPF neutrino detectors will only measure an event rate --- the convolution of the flux and the cross section --- so to make measurements of either, one must make assumptions about the other. This work introduces the possibility of using the \lownu method as a way to break that degeneracy, by providing a way to constrain the \numu flux shape at the FPF. We show that the impact of the DIS modeling and final-state interactions is small, and that the method is relatively model-independent for \numu. The impact is larger for the \numub flux, with some model-dependence seen. However, the method may still prove useful as a cross-check of the \numub flux shape, albeit with more caution required. Of particular importance is the quantification of the contribution from higher $q_{0}/E_\nu$ terms to the \lownu samples selected, for which data-driven constraints may be possible with a more sophisticated method.

We have demonstrated that the low-\qz portion of the charged current \numu--\argon and \numu--\tungsten cross sections, $\qz \leq 20$ GeV, is almost constant as a function of \enutrue, and that the small \enutrue-dependent corrections are consistent for two different neutrino interaction models. The  $\qz \leq 10$ GeV portion of the charged current \numub--\argon and \numub--\tungsten cross sections is also relatively constant with \enutrue, although larger model-dependent differences were observed. By making relatively simple detector response assumptions for both the \faser and FLArE FPF subdetector designs, we show that a low-\qz sample can be relatively well selected, although feed-down from higher-\qz introduces some additional model-dependence, particularly for the \numub case. We have shown that the event rates for these \lownu candidate samples, around 1\% of the total charged-current inclusive event rate for the fluxes and targets discussed, and that these samples can be used to constrain the \numu (\numub) flux-shape uncertainty to 5--10\% (10--20\%) bin-to-bin. 
Finally, we prove that these constraints are sufficient to discriminate between realistic flux predictions produced with current light and charmed hadron production models.

We hope that this preliminary assessment of the \lownu method in the context of the FPF motivates the FPF community to develop a more complete analysis. By breaking the degeneracy in the neutrino cross section and flux, the \lownu method has the potential to expand the range of reliable and model-independent FPF analyses that will be possible.

\FloatBarrier
\section*{Acknowledgments}
We acknowledge help from Felix Kling and Max Fieg in understanding the flux predictions for the various FPF detectors, and useful discussions with Aki Ariga on the detector assumptions used. We also thank Felix Kling for helpful comments on the manuscript.
AG acknowledges support from the European Union’s H2020-MSCA Grant Agreement No.101025085. The work of CW was supported by the U.S. Department of Energy, Office of Science, Office of High Energy Physics, under contract number DE-AC02-05CH11231.
This research used resources of the National Energy Research Scientific Computing Center (NERSC), a U.S. Department of Energy Office of Science User Facility located at Lawrence Berkeley National Laboratory, operated under Contract No. DE-AC02-05CH11231 using NERSC award HEP-ERCAP0023704.

\bibliographystyle{apsrev4-1}
\bibliography{lownu}

\end{document}